\documentclass[sigconf]{acmart}
\AtBeginDocument{%
  \providecommand\BibTeX{{%
    \normalfont B\kern-0.5em{\scshape i\kern-0.25em b}\kern-0.8em\TeX}}}
\usepackage{graphicx,tabularx,subcaption}
\usepackage{booktabs}
\usepackage{multirow}
\usepackage[capitalise,nameinlink,compress]{cleveref}
\usepackage{colortbl}
\usepackage{xcolor}
\usepackage{comment}
\usepackage[english]{babel}
\usepackage{fontawesome5}
\usepackage{arydshln}
\usepackage{tikz}
\usepackage{rotating}
\usepackage{enumitem}
\graphicspath{{images/}}
\usepackage{framed}
\usepackage{array}

\newenvironment{frshaded*}{%
\MakeFramed {\advance\hsize-\width \FrameRestore}
}%
{\endMakeFramed}

\newcommand*\rot[1][90]{\rotatebox{#1}}

\definecolor{main}{HTML}{fd829a}    
\definecolor{sub}{HTML}{fee2e8}     

\definecolor{lightgray}{gray}{0.9}

\definecolor{greenShade}{RGB}{239, 255, 232}
\definecolor{greenFont}{RGB}{34, 139, 34}

\definecolor{pinkShade}{RGB}{254, 236, 240}
\definecolor{pinkFont}{RGB}{251, 29, 71}

\definecolor{lightBlueShade}{RGB}{217, 243, 255}
\definecolor{lightBlueFont}{RGB}{0, 122, 176}

\definecolor{peachShade}{RGB}{255, 231, 209}
\definecolor{peachFont}{RGB}{242, 115, 0}

\definecolor{blueShade}{RGB}{231, 239, 253}
\definecolor{blueFont}{RGB}{19, 99, 223}

\definecolor{purpleShade}{RGB}{241, 232, 248}
\definecolor{purpleFont}{RGB}{112, 48, 160}

\definecolor{yellowShade}{RGB}{255, 253, 235}
\definecolor{yellowFont}{RGB}{204, 160, 29}

\definecolor{orangeShade}{RGB}{252, 230, 208}
\definecolor{orangeFont}{RGB}{249, 111, 7}

\copyrightyear{2025} 
\acmYear{2025} 
\setcopyright{acmlicensed}\acmConference[CHI '25]{CHI Conference on Human Factors in Computing Systems}{April 26-May 1, 2025}{Yokohama, Japan}
\acmBooktitle{CHI Conference on Human Factors in Computing Systems (CHI '25), April 26-May 1, 2025, Yokohama, Japan}
\acmDOI{10.1145/3706598.3713497}
\acmISBN{979-8-4007-1394-1/25/04}

\begin{document}


\title[Explanatory Debiasing]{Explanatory Debiasing: Involving Domain Experts in the Data Generation Process to Mitigate Representation Bias in AI Systems}

\author{Aditya Bhattacharya}
\orcid{0000-0003-2740-039X}
\email{aditya.bhattacharya@kuleuven.be}
\affiliation{%
  \institution{KU Leuven}
  \city{Leuven}
  \country{Belgium}
}

\author{Simone Stumpf}
\orcid{0000-0001-6482-1973}
\email{Simone.Stumpf@glasgow.ac.uk}
\affiliation{%
  \institution{University of Glasgow}
  \city{Glasgow}
  \country{Scotland, UK}
}

\author{Robin De Croon}
\orcid{0000-0002-1329-156X}
\email{robin.decroon@kuleuven.be}
\affiliation{%
  \institution{KU Leuven}
  \city{Leuven}
  \country{Belgium}
}

\author{Katrien Verbert}
\orcid{0000-0001-6699-7710}
\email{katrien.verbert@kuleuven.be}
\affiliation{%
  \institution{KU Leuven}
  \city{Leuven}
  \country{Belgium}
}

\renewcommand{\shortauthors}{Bhattacharya, et al.}

\begin{abstract}
Representation bias is one of the most common types of biases in artificial intelligence (AI) systems, causing AI  models to perform poorly on underrepresented data segments. Although AI practitioners use various methods to reduce representation bias, their effectiveness is often constrained by insufficient domain knowledge in the debiasing process. To address this gap, this paper introduces a set of generic design guidelines for effectively involving domain experts in representation debiasing.  We instantiated our proposed guidelines in a healthcare-focused application and evaluated them through a comprehensive mixed-methods user study with 35 healthcare experts. Our findings show that involving domain experts can reduce representation bias without compromising model accuracy. Based on our findings, we also offer recommendations for developers to build robust debiasing systems guided by our generic design guidelines, ensuring more effective inclusion of domain experts in the debiasing process.
\end{abstract}




\begin{teaserfigure}
  \centering
  \includegraphics[width=1.0\linewidth]{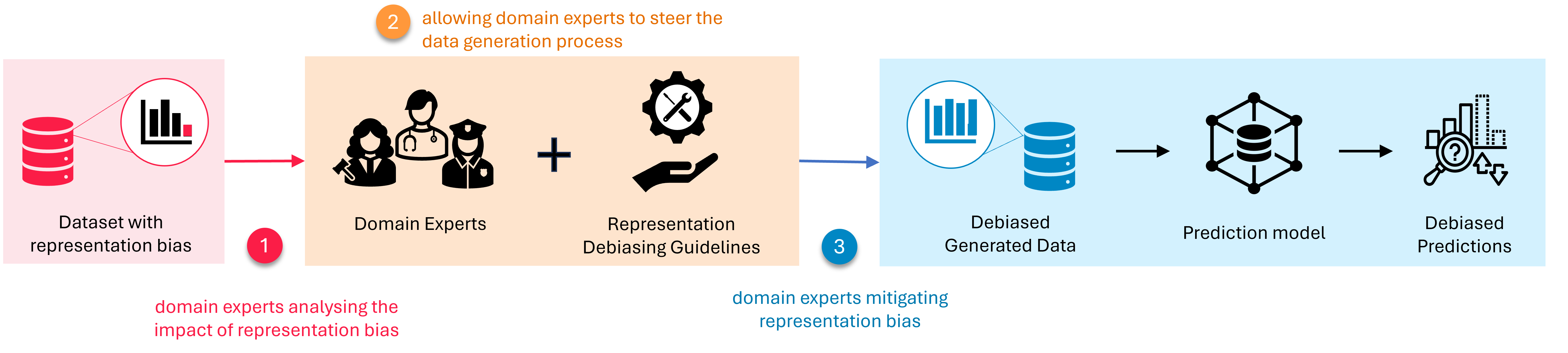}
  \caption{Explanatory debiasing involves empowering domain experts to understand representation bias and steer data generation algorithms for addressing its impact on AI systems. This paper introduces our generic design guidelines for involving domain experts in representation debiasing.}
  \Description[Explanatory debiasing]{Explanatory debiasing involves empowering domain experts to understand representation bias and steer data generation algorithms for addressing representation bias. This paper introduces our generic design guidelines for involving domain experts in representation debiasing}
  \label{fig:teaser_image}
\end{teaserfigure}

\maketitle

\section{Introduction}
Despite the significance of Artificial Intelligence (AI) and Machine Learning (ML) across various sectors, AI/ML systems are known to be affected by a wide spectrum of biases \cite{mehrabi2022survey}. One of the most prevalent types of biases in AI/ML systems that do not follow a systematic data collection process is \textit{representation bias} \cite{Shahbazi2023}. Representation bias arises when the training data lacks sufficient samples from certain groups or sub-groups \cite{Shahbazi2023,mehrabi2022survey}. Consequently, prediction models built on such a dataset will be biased towards other groups that have a higher representation. For example, if a healthcare dataset predominantly contains images of skin conditions from lighter-skinned individuals and has very few images from darker-skinned individuals, this creates representation bias. Consequently, an AI model trained for diagnosing skin conditions will perform poorly for individuals with darker skin tones \cite{BENCEVIC2024108044}. Due to the limited availability of data for under-represented populations, the model struggles to generalise effectively to these groups.

\textit{Data augmentation} is considered one of the most effective methods to reduce the impact of representation bias \cite{Iosifidis2018DealingWB, Shahbazi2023, sharma2020data}. It involves generating new synthetic data by applying various transformations to the existing data. Data augmentation is particularly useful when collecting additional data for under-represented groups is not practically feasible \cite{Shahbazi2023, dataaugment2023}. Adding new synthetic samples through data augmentation increases the representativeness of the underrepresented groups, thereby reducing the model’s bias towards the majority groups \cite{Shahbazi2023}.

However, despite increasing the number of samples, data augmentation algorithms have been criticised for generating problematic data points \cite{alkhawaldeh2023challenges, bareła2023dataaugment, balestriero2022effects, superb-ai-blog}. For instance, suppose there is a diabetes prediction dataset with predictor variables such as blood sugar level, body mass index (BMI), age and obesity level of patients. Data augmentation algorithms can generate records where the obesity level is high while the BMI is less than 18. However, in reality, a BMI under 18 indicates that a patient is underweight, not obese. Moreover, data augmentation algorithms such as SMOTE \cite{Chawla_2002}, ADASYN \cite{ADASYN2008}, CTGAN \cite{Gan2024}, TVAE \cite{VAE2014} have faced scrutiny for introducing data issues such as data drift, correlated features, duplicates, or outliers. Furthermore, since these augmentation algorithms tend to preserve the statistical properties of the training data, any form of bias that exists in the underlying training data may also exist in the generated data \cite{MUMUNI_DAC, balestriero2022effects, superb-ai-blog}. 

To mitigate these limitations, prior works have suggested involving domain experts \cite{ratner2017, rahman2020amplifying} such as healthcare professionals, legal advisors and policy makers. Domain experts are users who might not have AI/ML knowledge but are considered to be experts in their respective application domains. It has been argued in recent works \cite{bhattacharya2024exmos, feuerriegel2020fair, bhattacharya2023_technical_report} that the domain knowledge possessed by domain experts is crucial for identifying and modifying problematic data points and improving prediction models. However, due to the lack of user studies involving domain experts focused on mitigating representation bias in AI/ML systems, a significant gap exists in the literature regarding the potential involvement of domain experts in the representation debiasing process.

Our research addresses this gap by exploring how domain experts can be integrated into the data augmentation process to mitigate representation bias. In this paper, we propose a set of generic design guidelines for a representation debiasing system that involves domain experts. We instantiated these guidelines into a healthcare-focused prototype system. Moreover, we conducted a mixed-methods user study with 35 healthcare experts to investigate the following research questions:

\begin{description}[topsep=0pt, itemsep=0pt]
\item[RQ1.] How does involving domain experts during representation debiasing affect the prediction model’s performance?

\item[RQ2.] How can domain experts contribute to reducing representation bias in AI/ML systems?

\item[RQ3.] How does the involvement of domain experts in the representation debiasing process influence their trust in AI systems? How could the debiasing process be improved? 
\end{description}

\noindent The primary objective of this study is not to establish that domain experts are superior to AI experts in representation debiasing. Instead, our work emphasises the value of incorporating domain experts into the debiasing process, as our findings demonstrate that their involvement can effectively reduce representation bias in AI systems without compromising model performance. Therefore, domain experts can be essential contributors to the representation debiasing process, working alongside AI experts to ensure balanced and unbiased AI outcomes. To summarise, the contributions of this research are three-fold:
\begin{enumerate}[topsep=0pt, itemsep=0pt]
    \item  \textbf{Theoretical Contributions}: We propose a set of generic design guidelines for including domain experts in the representation debiasing process. While these guidelines are designed to be broadly applicable across various domains, we specifically evaluated their effectiveness in the healthcare domain.
    \item  \textbf{Artifact Contributions}: Based on our generic design guidelines, we developed a healthcare application following an iterative user-centric design process \cite{UCD2014}. This application supports healthcare experts in identifying predictor variables with representation bias. The application also allows them to share their domain knowledge during data augmentation to mitigate representation bias. The source code, design and architecture of the debiasing system is open-sourced on \href{https://github.com/adib0073/explanatory_debiasing}{GitHub}.
    \item  \textbf{Empirical Contributions}: To evaluate the healthcare-focused debiasing system developed using our design guidelines, we conducted a mixed-methods user study with 35 healthcare experts. Through the findings from this extensive user study, we show that domain experts can be involved in the representation debiasing process to reduce representation bias without compromising the prediction model's accuracy. Additionally, we found that including them in the debiasing process increased their overall trust in the AI system. Furthermore, we discuss how their involvement can be improved and how they can support AI experts in mitigating the impact of representation bias.
\end{enumerate}

\section{Background and Related Work}
\subsection{Representation Bias}
Representation bias arises when datasets lack sufficient information for various subgroups, often due to historical trends, skewed data distributions, flawed data preparation and acquisition methods, and selection and sampling biases when preparing the training dataset \cite{Shahbazi2023, mehrabi2022survey}. It causes a significant challenge for achieving group or sub-group fairness, as models trained on underrepresented groups tend to have lower accuracy for these groups \cite{Shahbazi2023}. To address this problem, training datasets must include adequate samples from less represented segments to improve prediction model performance \cite{Shahbazi2023}. 

To estimate the amount of representation bias, prior works have suggested two main metrics: (1) representation rate and (2) data coverage \cite{ Shahbazi2023, mousavi2023datacoverage}. For a specific variable, the \textit{representation rate} of a sub-group is measured by taking the ratio of its number of samples to the maximum number of samples among all sub-groups for that variable. For example, if a certain variable has three sub-groups: $a$, $b$ and $c$, such that $a + b + c = N$, where $N$ is the total number of samples, the representation rate of $a$ ($r_a$) is estimated by the equation: $r_a = \frac{a}{max(a, b, c)}$. \textit{Data coverage} refers to the minimum number of samples required for each sub-category to ensure its comprehensive representation within the dataset. The data coverage threshold is usually obtained after performing model over-fitting or under-fitting analysis \cite{Shahbazi2023}. Regardless of the dataset or the application domain, ensuring sufficient coverage for all significant sub-populations is crucial to guarantee their adequate representation.

Now, let us take an example to understand these metrics better. Consider a variable representing the severity of a medical condition in a dataset with three sub-categories as severity levels: mild, moderate, and severe. Assume that the dataset includes 900 patient records, with the following distribution of severity levels: 500 patients with mild severity, 150 with moderate severity, and 250 with severe severity. The representation rate for each severity level can be calculated as follows: $r_{mild} = \frac{500}{max(500, 150, 250)} = 1.0$, $r_{moderate} = \frac{150}{max(500, 150, 250)} = 0.3$ and $r_{severe} = \frac{250}{max(500, 150, 250)} = 0.5$. In this example, the representation rate for patients with moderate severity is 0.30, indicating that they are the least represented compared to patients with mild severity, who have a representation rate of 1.00, and those with severe severity, who have a representation rate of 0.50.  Suppose the minimum data coverage for this medical dataset is set as 200; then the moderate severity sub-category does not meet the coverage criteria as it has only 150 samples.

\subsection{Data Augmentation}

While acquiring more data for under-represented segments using a systematic collection process is recommended to address representation bias, it is often practically infeasible to collect additional data. Therefore, AI practitioners have proposed generating synthetic data as a viable alternative for representation debiasing \cite{Shahbazi2023, dataaugment2023, debiasaugment}. Since data augmentation involves generating new samples through various transformations on the underlying training data, it is seen as a potential solution for mitigating representation bias \cite{dataaugment2023, Shahbazi2023, bareła2023dataaugment}. 

For tabular datasets,  algorithms such as SMOTE \cite{Chawla_2002}  and ADASYN \cite{ADASYN2008}  have been predominantly used to generate synthetic data. More recently, generative AI algorithms such as CTGAN and TVAE \cite{ctgan2019} have also been considered for generating synthetic data from the existing training data. However, these methods have been criticised for introducing data issues such as outliers, redundant samples, and data drifts \cite{limitationsmote2022, alkhawaldeh2023challenges, balestriero2022effects}. 
Additionally, as they aim to maintain the statistical properties of the original data \cite{MUMUNI_DAC, balestriero2022effects, superb-ai-blog}, these methods often preserve the existing biases of the underlying training data in the generated samples. 

For instance, consider an underrepresented dataset for predicting heart disease that includes factors such as cholesterol levels, blood pressure, and age. Commonly adopted data augmentation algorithms can generate samples where the cholesterol level is high, but the blood pressure is unusually low. In practice, individuals with high cholesterol levels typically have elevated blood pressure. This inconsistency indicates that the augmented data may not accurately reflect real-world observations. Therefore, despite their success in generating synthetic data quickly, the lack of domain knowledge in the data augmentation process has been considered a challenge in producing synthetic data that accurately reflects real-world observations \cite{dataaugment2023, tang2020, temraz2021solving}. To produce accurate real-world representations of the generated data, our approach is to directly involve domain experts in the synthetic data generation process.

Our work highlights the crucial role of involving domain experts in the data generation process to mitigate representation bias. Leveraging the prior knowledge of domain experts can improve the creation of validated samples that accurately reflect real-world conditions, leading to more effective debiasing.

\subsection{Domain Expert Involvement in Refining Prediction Models}

Interactive machine learning (IML) researchers have often emphasised the importance of involving domain experts in model development, fine-tuning, and debugging \cite{fails2003, kulesza_explanatory_2010, kulesza_principles_2015, teso_leveraging_2022, teso2019, bhattacharya2024exmos, Schramowski2020, lakkaraju2022rethinking, Slack2023}. Different types of user interaction approaches have also been explored in prior research to enable the active involvement of domain experts. Their active involvement has been particularly helpful in identifying different types of biases and errors in the data \cite{2019DataQA, feuerriegel2020fair}. For example, prior knowledge possessed by healthcare experts has been considered essential for analysing and identifying biases and data issues in patient records that can lead to biased models \cite{ ueda2024fairness,  bhattacharya2024exmos}.

While prior research has introduced innovative interaction methods for domain experts to refine prediction models \cite{bhattacharya2024exmos, teso_leveraging_2022, teso2019, Schramowski2020}, there has been a notable lack of focus on their role in the data augmentation process for mitigating representation bias. Especially due to the lack of user studies on this topic, the perspective of domain experts is yet to be investigated. However, utilising the prior knowledge of domain experts holds considerable promise for reducing the risks associated with data augmentation algorithms used in tackling representation bias \cite{tang2020}. Our research addresses this gap by investigating how domain experts can contribute to the representation debiasing of generated data.

\section{Design Guidelines for Involving Domain Experts in Representation Debiasing}\label{sec_guidelines}
This section presents a set of generic design guidelines for effectively involving domain experts in data augmentation to mitigate representation bias. \Cref{tab:guidelines_summary} summarises these design guidelines, and an implementation of these guidelines in a healthcare scenario is presented in \Cref{sec_usage_scenario}. These guidelines are organised into three key phases of the debiasing process: pre-augmentation, during augmentation, and post-augmentation. By involving domain experts in the pre-augmentation phase, we ensure that the initial evaluation of the training dataset accurately identifies areas of underrepresentation. During the augmentation phase, their expertise is crucial in guiding the generation of synthetic data that faithfully reflects the characteristics of the underrepresented sub-categories. Finally, in the post-augmentation phase, domain experts play a vital role in validating and integrating the synthetic data with the existing dataset, ensuring that the model retraining process results in a more balanced and representative model. The methodology followed to conceptualise these guidelines is described in \Cref{sec_guideline_methods}.

\begin{table*}
\centering  
\caption{Summary of the design guidelines for involving domain experts in representation debiasing. This table presents the description and main purpose of each guideline elaborated in \Cref{sec_guidelines}.}
\label{tab:guidelines_summary}
\scalebox{.7}{
\begin{tabular}{@{}p{4cm}p{5cm}p{7cm}p{7cm}@{}}
\toprule
\textbf{Phase}                                & \textbf{Guidelines}                                               & \textbf{Description}                                                                                                                                                                                      & \textbf{Purpose}                                                                                                                                                                       \\ \midrule
\multirow{3}{*}{\parbox{3.8cm}{\colorbox{pinkShade}{ \textcolor{pinkFont}{\textbf{Pre-Augmentation }}} \\ Involving domain experts to identify biased variables and their impact on the model performance before the augmentation process}}     & \textsc{Exploration Through Data-Centric Explanations} & Provide data-centric explanations that help domain experts explore the representativeness of different segments for each predictor variable.                                                 & Help domain experts identify predictor variables with representation bias in the current training data before the augmentation algorithm is applied.                \\
\cmidrule(r){2-4}
& \textsc{Model Impact Analysis}                                   & Show the model performance across all segments for each predictor variable.                                                                                                                    & Help domain experts identify underrepresented segments where the model performance is the lowest so that these segments can be emphasised during the augmentation process \\
\cmidrule(r){2-4}
& \textsc{AI System Transparency}                                  & Show the overall model performance, number of training samples, predictor variables and the estimated quality of the training dataset.                                                       & Enable domain experts to assess the system's reliability, scope, and limitations, fostering a deeper understanding of its functionality and areas for improvement.        \\ \midrule
\multirow{2}{*}{\parbox{3.8cm}{\colorbox{peachShade}{ \textcolor{peachFont}{\textbf{During Augmentation }}} \\
Allowing domain experts to steer augmentation algorithms
}} & \textsc{Multivariate Constraint Planning}                      & Allow domain experts to select multiple predictor variables and their corresponding sub-segments that particularly need to be represented in the generated data.                            & Enable domain experts to steer data augmentation algorithms to generate practically plausible and more representative samples.                                            \\
\cmidrule(r){2-4}
& \textsc{Awareness of extremely low coverage segments}     & Show a warning when data segments with extremely low coverage are selected during the augmentation process.                                                                                  & Warn domain experts about the potential generation of problematic samples when segments with extremely low coverage are selected for data augmentation.                   \\ \midrule
\multirow{4}{*}{\parbox{3.8cm}{\colorbox{lightBlueShade}{ \textcolor{lightBlueFont}{\textbf{Post-Augmentation }}} \\
Involving domain experts in the validation and refinement of the generated data
}}   & \textsc{Refinement of Generated Data}                          & Provide control to domain experts for selecting only plausible synthetic samples, filtering out noisy data and modifying generated samples to correctly represent real-world observations. & Help domain experts in removing noisy and implausible generated data.                                                                                                       \\
\cmidrule(r){2-4}
& \textsc{Local What-If Exploration}                              & Allow domain experts to conduct \textit{what-if analysis} of the generated samples.                                                                                                                   & Help domain experts identify and rectify problematic samples from the generated dataset.                                                                                    \\
\cmidrule(r){2-4}
& \textsc{Evaluation of Generated  Data}                      & Apply the current prediction model to the generated data by considering the generated dataset as the inference set.                                                                          & Enhances the transparency of the generated data by presenting an estimate of the model performance and quality of the generated.                                          \\
\cmidrule(r){2-4}
& \textsc{User-Interaction Bias Awareness}                       & Inform domain experts about potential user interaction biases before retraining the system with the new training set that combines the original dataset with the generated dataset.          & Minimise the introduction of user interaction biases in the generated dataset.                                                                                              \\ 
\bottomrule
\end{tabular}
}
\end{table*}

\subsection{Pre-Augmentation}
To enable more effective feedback from domain experts during the data augmentation process, the following pre-augmentation guidelines are essential. These guidelines ensure that domain experts can accurately identify variables with representation bias and understand the impact of this bias on the prediction model's performance across underrepresented segments.:
\begin{enumerate}[start=1,label={\textbf{ \arabic*.}}, left=0cm]
    \item \colorbox{pinkShade}{ \textcolor{pinkFont}{\textsc{ Exploration Through Data-Centric Explanations}}}: Data-centric explanations are explainable AI (XAI) methods that focus on understanding and interpreting the training data itself rather than the models built from it \cite{BhattacharyaXAI2022, anik_data-centric_2021, Bhattacharya2023}. These explanations highlight the underlying patterns, distributions, and characteristics of the data and help domain experts identify the presence of representation bias and erroneous data points. To ensure usability and enhance the experience of domain experts, these explanations should be presented through intuitive and interactive interfaces that prioritise clarity, accessibility, and ease of exploration \cite{bove_contextualization_2022}. Interactive data-centric explanations further empower them to explore each predictor variable \cite{Bhattacharya2023} and identify segments with representation bias, which is crucial for effective data augmentation and bias mitigation.
    \item \colorbox{pinkShade}{ \textcolor{pinkFont}{\textsc{Model Impact Analysis}}}: To help domain experts understand the impact of underrepresented segments on the prediction model, model performance metrics such as accuracy, precision, and recall should be clearly demonstrated across all data segments, with a particular emphasis on underrepresented groups. By identifying segments of each predictor variable where model performance is significantly lower, domain experts can prioritise these segments for the augmentation process. This allows them to address potential representation bias that may hinder the model's ability to make informed decisions for such underrepresented groups.
    \item \colorbox{pinkShade}{ \textcolor{pinkFont}{\textsc{AI System Transparency}}}: Aligned with the \textit{ML transparency} principle from Bove et al. \cite{bove_contextualization_2022},  the overall AI system performance metrics, such as model accuracy, number of training samples, predictor variables, and estimation of dataset quality \cite{bhattacharya2024exmos}, should be clearly presented to domain experts through interactive visualisations and UI elements. Transparency about this system information allows domain experts to gauge the system’s reliability, scope, and potential limitations, thereby facilitating a deeper understanding of how the AI system functions and where it may require further refinement \cite{bellotti2001intelligibility}.  

\end{enumerate}

\subsection{During Augmentation}
To generate more meaningful synthetic data that correctly reflects real-world observations, the following guidelines can be applied to enhance the involvement of domain experts in the augmentation process. These guidelines ensure that data augmentation algorithms are steered by domain knowledge, creating synthetic data that closely aligns with real-world representations.
\begin{enumerate}[start=1,label={\textbf{ \arabic*.}}, left=0cm]
    \item \colorbox{peachShade}{ \textcolor{peachFont}{\textsc{Multivariate Constraint Planning}}}: Since data augmentation algorithms can generate samples with random and implausible values \cite{alkhawaldeh2023challenges}, prior research has recommended setting constraints on the training data for obtaining more meaningful data \cite{gao2023, MUMUNI_DAC}. According to this guideline, domain experts should be given control over selecting predictor variables and their corresponding segments that need better representation through interactive UI elements rather than applying data augmentation randomly across all segments. Additionally, domain experts may possess better in-depth knowledge of the training data to understand the joint impact of multiple predictor variables. By allowing them to define specific value ranges and the number of samples required for multiple predictor variables with representation bias, data augmentation algorithms can be steered to generate practically plausible and more representative samples.
    \item \colorbox{peachShade}{ \textcolor{peachFont}{\textsc{Awareness of extremely low coverage segments}}}: When sub-groups have very low representation, data augmentation algorithms may struggle to generate accurate synthetic samples \cite{alkhawaldeh2023challenges}. Therefore, significantly underrepresented data segments should be clearly highlighted through visual elements to assist domain experts in determining the necessary number of samples that should be generated for these segments. If the required number of synthetic samples greatly exceeds the sample size of the segment in the underlying training data, a warning should be issued to domain experts, as these synthetic samples are more likely to contain problematic data points.

\end{enumerate}
\subsection{Post-Augmentation}
The following post-augmentation guidelines ensure that the generated dataset is thoroughly scrutinised by domain experts, minimising the presence of problematic samples before it is integrated into the original training set and the model is retrained.
\begin{enumerate}[start=1,label={\textbf{ \arabic*.}}, left=0cm]
    \item \colorbox{lightBlueShade}{ \textcolor{lightBlueFont}{\textsc{Refinement of Generated Data}}}: Domain experts should be given the control to refine the generated data by selecting only plausible synthetic samples, particularly for underrepresented segments. To minimise the introduction of problematic data points during the debiasing process, we recommend providing data filters that enable sampling based on conditions specified by domain experts. Any interface designed to instantiate this guideline should provide visual cues and clear feedback to enhance the overall user experience during conditional sampling of the generated data and allowing users to filter out noisy data.
    \item \colorbox{lightBlueShade}{ \textcolor{lightBlueFont}{\textsc{Local What-If Exploration}}}: Domain experts should be given the ability to conduct local ``what-if`` exploration of generated samples \cite{Lim_CHI_2009, Bhattacharya2023, wang_designing_2019, BhattacharyaXAI2022}. \textit{Local what-If exploration} is a process that allows users to assess and manipulate individual data points within a dataset to observe how changes in predictor variables affect model outcomes. This interactive approach helps domain experts to validate the accuracy and reliability of model predictions for generated synthetic samples, enabling them to identify and rectify any problematic values. It provides more granular control to domain experts to edit or remove samples that are corrupt or practically implausible.
    
    \item \colorbox{lightBlueShade}{ \textcolor{lightBlueFont}{\textsc{Evaluation of Generated Data}}}: With this guideline, the current ML model's performance should be evaluated on the generated data before retraining with the augmented dataset, treating the generated set as an inference set. However, the generated data should be solely considered for augmenting the training set, with the final evaluation of the system being conducted on a separate and untouched dataset to avoid data leakage \cite{kapoor2022leakagereproducibilitycrisismlbased}. The main intention of this guideline is to provide an initial estimate of the quality and performance of the current model on the newly generated set. Additionally, we suggest assessing the quality of the generated data using a method similar to Bhattacharya et al. \cite{bhattacharya2024exmos} and displaying the model's confidence level for each synthetic sample, following Brownlee's method \cite{brownlee2019confidence}. This guideline enhances the transparency of the generated data, helping domain experts identify problematic samples to address representation bias before integrating with the existing training set.
    
    \item \colorbox{lightBlueShade}{ \textcolor{lightBlueFont}{\textsc{User-Interaction Bias Awareness}}}: Before retraining the system with the augmented dataset, domain experts should be informed about potential biases that can be introduced through their interactions with the system when performing data augmentation \cite{mehrabi2022survey}. To elucidate the user-interaction biases and their effects on the training dataset when incorporating generated data, we suggest using visual data-centric explanations \cite{Bhattacharya2023}. Data-centric explanations can highlight issues in the new training set, serving as a cautionary note.  This allows domain experts to address any identified adverse effects and, if necessary, adjust or revert the augmentation process.
    
\end{enumerate}

\subsection{Conceptualisation of Design Guidelines}\label{sec_guideline_methods}
We developed this set of generic design guidelines for debiasing based on an extensive literature review of guidelines for involving domain experts in AI system refinement. This process involved synthesising insights from multiple research areas, including Bias and Fairness in AI systems, Data Augmentation, Human-AI Interaction, and Explainable AI, to ensure comprehensive coverage of relevant perspectives. We adopted a structured approach to conceptualise these guidelines. First, we reviewed existing literature to identify causes and potential strategies for mitigating representation bias in datasets and models. Next, we investigated the opportunities and limitations of widely used data augmentation algorithms in addressing representation bias, in which we identified the importance of incorporating domain expertise during the augmentation process. Subsequently, we explored works focusing on integrating domain knowledge into data augmentation processes, examining strategies for leveraging domain expertise at various stages of data augmentation. Finally, we analysed prior research on user interface (UI) design and interaction frameworks to gather insights into enhancing the experience of domain experts involved in debiasing workflows. We refined the guidelines iteratively, particularly after an exploratory feedback session with five healthcare experts (discussed later in \Cref{sec_exploratory_study}). \Cref{tab:principle_methods} highlights the literature that informed the conceptualisation of these design guidelines. 

\begin{table*}
\centering
\caption{Summary of existing work explored to conceptualise the design guidelines for domain expert involvement in representation debiasing. This table also highlights the key takeaways from prior works for the development of the design guidelines.}
\label{tab:principle_methods}

\begin{center}
\scalebox{0.75}{
\begin{tabular}[c]{c|>{\centering\arraybackslash}m{1.5cm}:>{\centering\arraybackslash}m{1.5cm}|>{\centering\arraybackslash}m{1cm}:>{\centering\arraybackslash}m{1cm}:>{\centering\arraybackslash}m{1cm}|>{\centering\arraybackslash}m{1.5cm}:>{\centering\arraybackslash}m{1.5cm}|>{\centering\arraybackslash}m{1cm}:>{\centering\arraybackslash}m{1cm}:>{\centering\arraybackslash}m{1cm}}

\cline{2-11}\noalign{\smallskip}
 & \multicolumn{2}{c|}{\textbf{\parbox{3cm}{\centering {Representation Bias\\in AI Systems }}}} & \multicolumn{3}{c|}{\textbf{\parbox{3cm}{\centering {Data Augmentation\\for Debiasing}}}} &
\multicolumn{2}{c|}{\textbf{\parbox{3cm}{\centering Domain Expertise\\in Augmentation}}} &
\multicolumn{3}{c}{\textbf{\parbox{3cm}{\centering {Designing Interaction Methods\\}}}} \\
\noalign{\smallskip}
\cline{2-11}\noalign{\smallskip}
& \rot{{\parbox{3cm}{\centering Causes and Effects}}} & \rot{{\parbox{3cm}{\centering Potential Mitigation Strategies}}} & \rot{{\parbox{3cm}{\centering Potential in Mitigating Representation Bias}}} & \rot{{\parbox{3cm}{\centering Pros and Cons of Different Techniques}}} & \rot{{\parbox{3cm}{\centering Need for Domain Knowledge}}} & \rot{{\parbox{3cm}{\centering Incorporating Domain Knowledge in\\Different Augmentation Stages}}} & \rot{{\parbox{3cm}{\centering Importance of Validation of Generated Data}}} & \rot{{\parbox{3cm}{\centering Domain Expert Involvement}}} & \rot{{\parbox{3cm}{\centering Improving System Understanding using XAI}}} &
\rot{{\parbox{3cm}{\centering UI Design Guidelines for Domain Expert Involvement}}}  \\
\hline
\rowcolor{gray!10}
\citeauthor{Shahbazi2023} (\citeyear{Shahbazi2023}) & $\bullet$ & $\bullet$ & $\bullet$ & & $\bullet$ & $\bullet$ & & & &\\ 
\citeauthor{mehrabi2022survey} (\citeyear{mehrabi2022survey}) & $\bullet$ & $\bullet$ & $\bullet$ & & & & & & &\\ 
\rowcolor{gray!10}
\citeauthor{mousavi2023datacoverage} (\citeyear{mousavi2023datacoverage}) & $\bullet$ & $\bullet$ & & & & & & & &\\ 
\citeauthor{BENCEVIC2024108044} (\citeyear{BENCEVIC2024108044}) & $\bullet$ & $\bullet$ &  & &  & & & & &\\ \rowcolor{gray!10}
\citeauthor{Iosifidis2018DealingWB} (\citeyear{Iosifidis2018DealingWB}) &  & $\bullet$ & & & & & & & &\\ 
\citeauthor{sharma2020data} (\citeyear{sharma2020data}) &  & $\bullet$ & $\bullet$ &  & $\bullet$ & $\bullet$ & & & &\\ \rowcolor{gray!10}
\citeauthor{dataaugment2023} (\citeyear{dataaugment2023}) &  &  & $\bullet$ & & $\bullet$ & $\bullet$ & $\bullet$ & & &\\ 
\citeauthor{alkhawaldeh2023challenges} (\citeyear{alkhawaldeh2023challenges}) &  &  & $\bullet$ & $\bullet$ & $\bullet$ & & $\bullet$ & & &\\ \rowcolor{gray!10}
\citeauthor{bareła2023dataaugment} (\citeyear{bareła2023dataaugment}) &  & $\bullet$ & $\bullet$ & & $\bullet$ & & & & $\bullet$ &\\ 
\citeauthor{balestriero2022effects} (\citeyear{balestriero2022effects}) &  &  & & $\bullet$ & $\bullet$ & $\bullet$ & & & &\\ \rowcolor{gray!10}
\citeauthor{UmapBhattacharya2024} (\citeyear{UmapBhattacharya2024}) & $\bullet$ & $\bullet$ & $\bullet$ & & $\bullet$ & & $\bullet$ & $\bullet$ & & $\bullet$\\ 
\citeauthor{Chawla_2002} (\citeyear{Chawla_2002}) &  &  & & $\bullet$ &  $\bullet$ & & & & &\\ \rowcolor{gray!10}
\citeauthor{ADASYN2008} (\citeyear{ADASYN2008}) &  &  & & $\bullet$ & $\bullet$ & & & & &\\ 
\citeauthor{ctgan2019} (\citeyear{ctgan2019}) &  & & & $\bullet$ & $\bullet$ &  & $\bullet$ & & &\\ \rowcolor{gray!10}
\citeauthor{MUMUNI_DAC} (\citeyear{MUMUNI_DAC}) &  &  & & $\bullet$ & $\bullet$ & & & & &\\ 
\citeauthor{ratner2017} (\citeyear{ratner2017}) & $\bullet$ & $\bullet$ & & & $\bullet$ & $\bullet$ & $\bullet$ & & &\\ \rowcolor{gray!10}
\citeauthor{rahman2020amplifying} (\citeyear{rahman2020amplifying}) & &  & & & $\bullet$ & & & $\bullet$ & &\\ 
\citeauthor{bhattacharya2024exmos} (\citeyear{bhattacharya2024exmos}) &  &  & & & & &  & $\bullet$ & $\bullet$ & $\bullet$\\ \rowcolor{gray!10}
\citeauthor{feuerriegel2020fair} (\citeyear{feuerriegel2020fair}) &  &  & & & & $\bullet$ & $\bullet$ &  &  &\\ 
\citeauthor{tang2020} (\citeyear{tang2020}) &  &  & & $\bullet$ & $\bullet$ & $\bullet$ & $\bullet$ & & &\\ \rowcolor{gray!10}
\citeauthor{temraz2021solving} (\citeyear{temraz2021solving}) &  & $\bullet$  & $\bullet$ & $\bullet$ & & & &  & $\bullet$ &\\ 
\citeauthor{kulesza_explanatory_2010} (\citeyear{kulesza_explanatory_2010}) &  &  & & & & & & $\bullet$ & $\bullet$ & $\bullet$\\ \rowcolor{gray!10}
\citeauthor{kulesza_principles_2015} (\citeyear{kulesza_principles_2015}) &  &  & & & & & & & $\bullet$ & $\bullet$\\ 
\citeauthor{teso_leveraging_2022} (\citeyear{teso_leveraging_2022}) &  & & & & & &$\bullet$ & &$\bullet$ &\\ \rowcolor{gray!10}
\citeauthor{Schramowski2020} (\citeyear{Schramowski2020}) &  & & & & & & & $\bullet$ & $\bullet$ &\\ 
\citeauthor{lakkaraju2022rethinking} (\citeyear{lakkaraju2022rethinking}) &  & & & & & & & $\bullet$ & $\bullet$ &\\ \rowcolor{gray!10}
\citeauthor{Slack2023} (\citeyear{Slack2023}) & &  & & & & & & $\bullet$ & $\bullet$ &\\ 
\citeauthor{teso2019} (\citeyear{teso2019}) &  &  & & & & & & $\bullet$ & $\bullet$ & $\bullet$\\ \rowcolor{gray!10}
\citeauthor{ueda2024fairness} (\citeyear{ueda2024fairness}) &  & $\bullet$ & $\bullet$ & & $\bullet$ & & $\bullet$ & & &\\ 
\citeauthor{BhattacharyaXAI2022} (\citeyear{BhattacharyaXAI2022}) &  &  & & & & & & $\bullet$ & $\bullet$ & $\bullet$\\ \rowcolor{gray!10}
\citeauthor{anik_data-centric_2021} (\citeyear{anik_data-centric_2021}) &  & & & & & & & & $\bullet$ & $\bullet$\\ 
\citeauthor{Bhattacharya2023} (\citeyear{Bhattacharya2023}) &  &  & & & & & & & $\bullet$ & $\bullet$\\
\rowcolor{gray!10}
\citeauthor{bove_contextualization_2022} (\citeyear{bove_contextualization_2022}) &  &  & & & & & & & $\bullet$ & $\bullet$ \\ 
\citeauthor{bellotti2001intelligibility} (\citeyear{bellotti2001intelligibility}) &  &  & & & & & & $\bullet$ & $\bullet$ &\\ 
\rowcolor{gray!10}
\citeauthor{Lim_CHI_2009} (\citeyear{Lim_CHI_2009}) &  &  & & & & & & & $\bullet$ & $\bullet$\\ 
\citeauthor{wang_designing_2019} (\citeyear{wang_designing_2019}) &  &  & & & & & & $\bullet$ & $\bullet$ & $\bullet$\\ 
\bottomrule
\end{tabular}
} 
\end{center}
\end{table*}

\section{Application: Implementation of Design Guidelines in a Healthcare Scenario}\label{sec_usage_scenario}
We applied the generic design guidelines outlined in \Cref{sec_guidelines} to develop a healthcare-focused application by following an iterative user-centric design process \cite{UCD2014}. This section details the application's usage scenario, implementation, and its various user interface (UI) components.

\subsection{Usage Scenario}
We applied our proposed design guidelines into a debiasing application tailored to include healthcare domain experts such as doctors, nurses, and medical researchers. The application features an ML model that predicts the onset of type 2 diabetes based on patient medical records. It enables healthcare experts to identify biased predictor variables in the training dataset and offer feedback during the data augmentation process. Additionally, it allows them to validate and refine the generated data before integrating it into the training set and retraining the model.

\subsection{Application Implementation}
\emph{Low Fidelity Prototype}\label{sec_exploratory_study}: We followed an iterative user-centric design process \cite{UCD2014} to implement our debiasing application. We first created an initial low-fidelity click-through prototype in Figma \cite{figmaUrl} to instantiate the generic design guidelines. Then, to refine this prototype, we conducted an exploratory feedback session that included voluntary pro-bono participation from 5 healthcare experts (2 females, 3 males; aged between 29 and 51 years; each having more than 4 years of healthcare experience) employed at the \anon{the Faculty of Healthcare Sciences, University of Maribor in Slovenia}. We obtained the ethical approval for this feedback session from \anon{KU Leuven} with an approval number \anon{G-2024-8352-R2(MAR)}. These feedback sessions consisted of individual co-design and think-aloud sessions that were recorded and later transcribed for analysis. The average duration for each session was around 30 minutes.

\noindent\emph{Key Takeaways from the Exploratory Session}: The purpose of this initial design feedback session was to validate our design guidelines and UI designs and ensure alignment with the primary user requirements for involving domain experts in the debiasing process. From participant feedback, we identified key improvements, such as simplifying the navigation between the exploration of representation bias in the predictor variables and setting constraints during the augmentation process to reduce cognitive load. Initially, we planned to separate variable exploration and constraint-setting for the data augmentation algorithm into two distinct views, similar to the EXMOS approach in Bhattacharya et al. \cite{bhattacharya2024exmos}. However, most of the participants suggested keeping these features in a single view to reduce their cognitive load. Additionally, we learned the importance of tooltips when describing each UI component and clearly highlighting biased sub-groups and their impact through clear warning messages from our participant feedback. These insights informed the development of a high-fidelity debiasing application tailored to the needs of healthcare experts to effectively engage them in the debiasing process.

\noindent\emph{Debiasing Application and ML Model}: The high-fidelity application was developed as an interactive web application using React.js, with a backend ML engine built in Python. It included a type-2 diabetes prediction model developed using the LightGBM classification algorithm \cite{LightGBM2017}. While our proposed guidelines for representation debiasing are model-agnostic (i.e., they do not depend on the specific ML algorithm used), we selected LightGBM for its efficiency, accuracy, and significantly lower memory consumption. The prediction model had an accuracy of 93\%. Additional information about our model training process is available in the supplementary material.

\noindent\emph{Dataset}: The model was trained on an open-sourced type-2 diabetes prediction dataset \cite{DiabetesDatasetPublication}. The dataset comprises 4,303 patient records, with 17 predictor variables and 1 target variable. Each record includes details on the patient's physical attributes (e.g., age, gender, BMI, etc.), diagnostic measures (e.g., plasma glucose,  cholesterol measures, etc.), and self-reported information such as smoking and drinking habits, as well as family history of diabetes. A complete list of predictor variables, along with their full descriptions, is provided in the supplementary material. We selected this dataset for our experiments because it was collected according to World Health Organization (WHO) standards, ensuring its reliability for developing diabetes diagnosis models. However, the dataset contains several predictor variables with representation bias, making it a suitable proxy for real-world clinical datasets with representation biases.

\noindent\emph{Data Augmentation Method}: For conducting data augmentation, we used the Conditional Tabular Generative Adversarial Network (CTGAN) algorithm \cite{ctgan2019}, a generative AI method recognised for generating better real-world tabular data compared to alternatives, like SMOTE \cite{Chawla_2002} and ADASYN \cite{ADASYN2008}. CTGAN leverages a Generative Adversarial Network (GAN) architecture with a generator and a discriminator working in tandem to learn and replicate the joint distribution of the input data. It incorporates mode-specific normalisation and a conditional sampling mechanism, enabling it to handle mixed data types (continuous, categorical, and discrete) while preserving complex feature relationships and addressing imbalanced datasets.  While our proposed principles are not tied to any specific augmentation algorithm, we chose CTGAN due to its fast computation speed and the relatively low incidence of data issues in the generated samples.

\begin{figure*}
\centering
\includegraphics[width=0.9\linewidth]{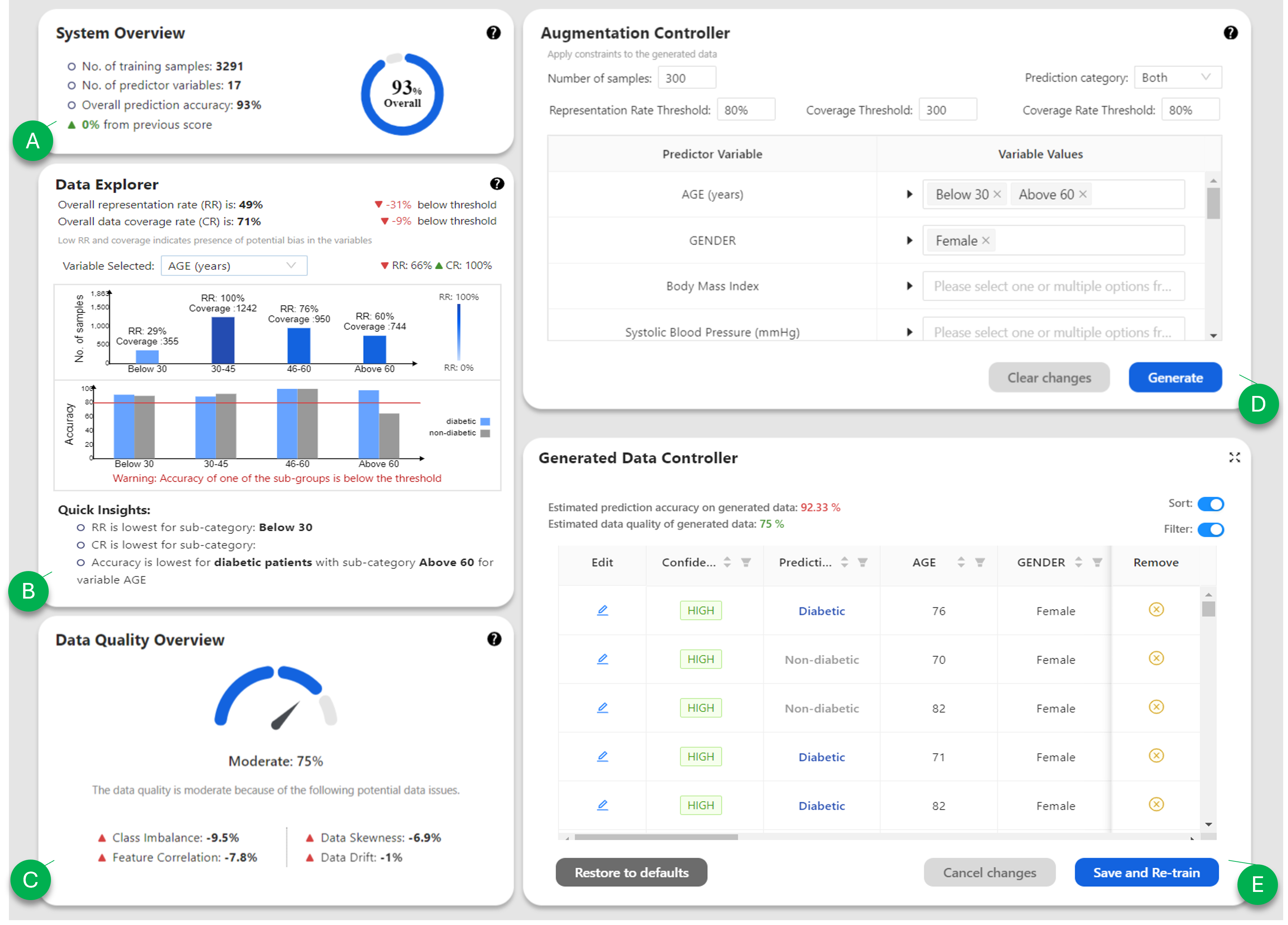}
\caption{Screenshot of our debiasing application showing the following UI components described in \Cref{subsec_UI_components}: (A) System Overview (B) Data Explorer (C) Data Quality Overview (D) Augmentation Controller (E) Generated Data Controller. }
\Description[UI for healthcare-focused debiasing application]{Screenshot of our debiasing application showing the following UI components described in \Cref{subsec_UI_components}: (A) System Overview (B) Data Explorer (C) Data Quality Overview (D) Augmentation Controller (E) Generated Data Controller.}
\label{fig:ui_components}
\end{figure*}

\begin{figure*}
\centering
\framebox{\includegraphics[width=0.55\linewidth]{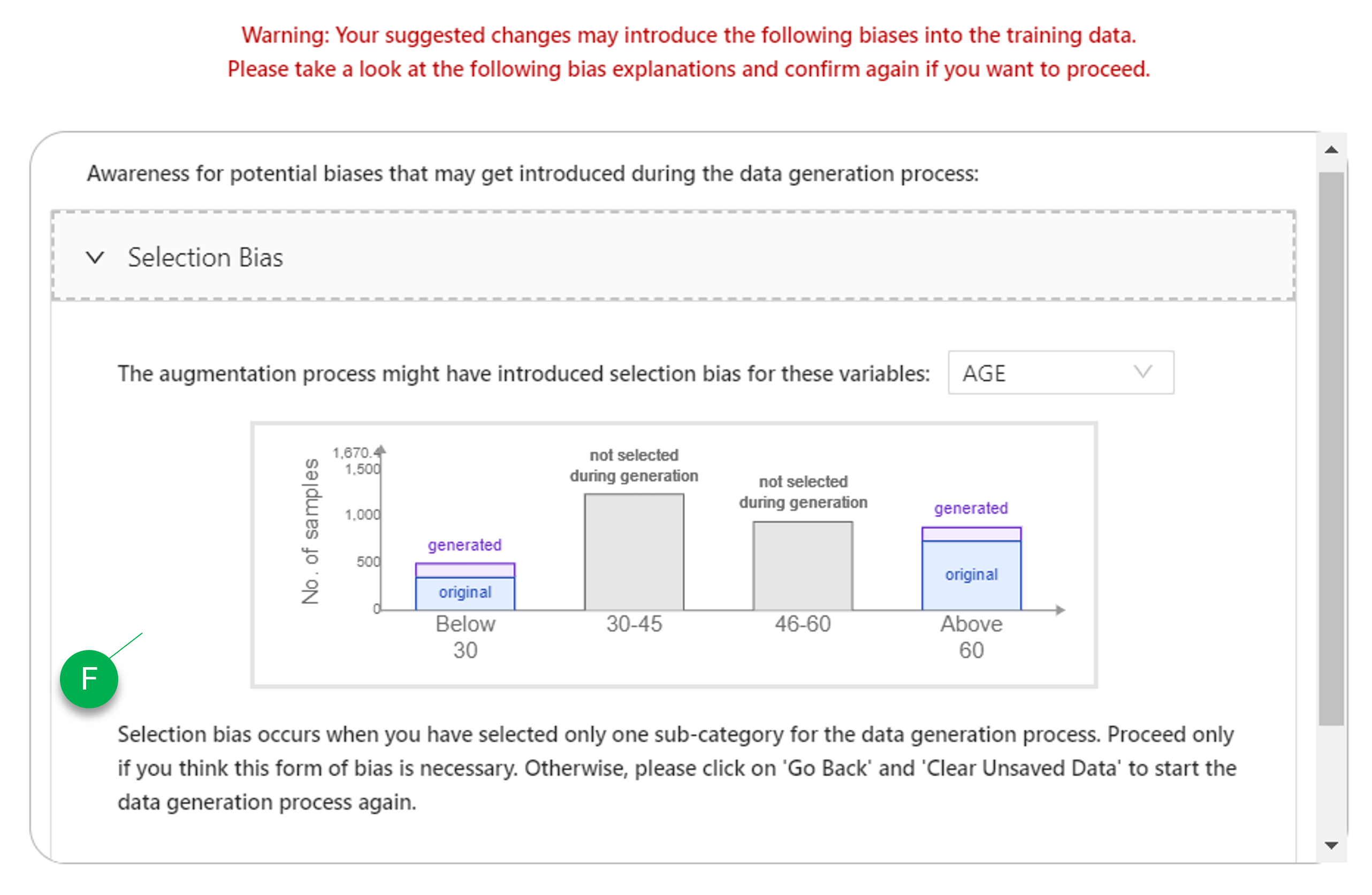}
}
\caption{Screenshot of the User-Interaction Bias Awareness component from \Cref{subsec_UI_components}., which issues a warning when users try to retrain the system with generated data. }
\Description[Screenshot of Bias Awareness UI component]{Screenshot of the User-Interaction Bias Awareness component from \Cref{subsec_UI_components}.}
\label{fig:bias_awareness}
\end{figure*}

\subsection{User Interface Components}\label{subsec_UI_components}

Based on the design guidelines discussed in \Cref{sec_guidelines}, our healthcare-focused application consists of the following UI components. \Cref{fig:ui_components} and \Cref{fig:bias_awareness} presents screenshots of our application illustrating these UI components.
\begin{enumerate}[start=1,label={\textbf{ (\Alph*)}}, left=0cm]
    
    \item \textbf{System Overview}: In line with the Pre-Augmentation \textsc{AI System Transparency} guideline, this UI component presents key information about the prediction model, including its overall accuracy, the number of training samples, and the predictor variables used. It also shows the change in accuracy if the model is retrained with the newly generated data. Additionally, tooltips are provided to explain the model's primary purpose and how it can assist healthcare experts. This component is most useful during the pre-augmentation phase and after the prediction model is retrained with the generated data, allowing users to compare changes to the overall prediction model due to the augmentation process.
    
    \item \textbf{Data Explorer}: Considering the \textsc{Exploration Through Data-Centric Explanations} guideline, this component allows healthcare experts to explore each predictor variable and pinpoint those that are impacted by representation bias. In addition to displaying the overall representation rate (RR) and coverage rate (CR) considering all the predictor variables, it also provides RR and CR scores for individual variables. Furthermore, considering the \textsc{Model Impact Analysis} guideline, this component displays not only the frequency distribution of each segment of the predictor variable but also their impact on the model accuracy for different target outcomes. This component also includes a \textit{Quick Insights} section, inspired by Bhattacharya et al.'s \cite{bhattacharya2024exmos} design suggestions of \textit{Key Insights}. This section highlights the segments of each predictor variable that are most affected, where RR, CR, and accuracy scores are notably low. Overall, this component is most useful during the pre-augmentation phase in which users get to explore the representation bias in predictor variables.
    
    \item \textbf{Data Quality Overview}: This UI component provides transparency about the overall quality of the current training dataset. It is aligned with the Pre-Augmentation \textsc{AI System Transparency} guideline. We followed the method prescribed by Bhattacharya et al. \cite{bhattacharya2024exmos} to compute the data quality. It includes the detection of common data issues such as outliers, redundant records, correlated features, skewed variables and class imbalance. Each data issue is given an equal weightage in the scoring process of estimating the overall data quality. The design of this component is also aligned with the guideline of showing \textit{uncertainty measures} from Wang et al. \cite{wang_designing_2019}.
    
    \item \textbf{Augmentation Controller}: Aligned with the Augmentation guidelines, this component allows healthcare experts to select predictor variables and segments that require representation in the generated data. It allows them to control the data augmentation process in order to mitigate the impact of representation bias. Additionally, a warning is issued if users try to generate a significantly higher number of samples for segments with extremely low representation in the training data.
    
    \item \textbf{Generated Data Controller}: After the augmentation process, newly generated data is displayed in this component, which implements the Post-Augmentation guidelines. It allows users to perform \textsc{Refinement of Generated Data} using various filters and sorting controls. Considering the \textsc{Evaluation of Generated Data} guideline, it shows the estimated accuracy and the quality of the generated data. The component also displays expected prediction outcomes and model confidence levels, helping users identify potentially faulty samples. It allows users to perform \textsc{Local What-If Exploration}, in which users can edit values of each generated sample to observe changes in predicted outcomes. If a sample is deemed irreparably faulty, it can be removed entirely. Thus, this component is designed primarily to allow users to validate and refine the generated data that is obtained after the augmentation algorithm is applied. 
    
    \item \textbf{User Interaction Bias Awareness}: After the validation process, users can proceed with retraining the model by incorporating the generated data.  However, a warning about potential user interaction biases, such as \textit{Selection Bias} and \textit{Generation Bias} \cite{mehrabi2022survey} is issued in accordance with the \textsc{User-Interaction Bias Awareness} guideline. Visual data-centric explanations \cite{Bhattacharya2023} are used to highlight how the interactions during the augmentation process may affect the existing training data. Users have the option to revert changes made to the generated data or even discard the generated dataset and redo the augmentation process. Although this component cautions users about potential biases, it does not restrict them from merging the generated data with the training set and retraining the model.

\end{enumerate}

\section{User Study}\label{sec_user_study}
This section discusses the methodology used to evaluate our final debiasing application, based on the generic design guidelines, through a mixed-methods user study involving 35 healthcare experts. The ethical approval of this study was granted by the ethical committee of \anon{KU Leuven} with the approval number \anon{G-2024-8352-R2(MAR)}.

\subsection{Study Setup}
We conducted a mixed-methods user study with 35 healthcare experts to investigate how domain experts, like healthcare experts, can be effectively involved in the representation debiasing process and seek answers to our research questions. The study was conducted online, with participants taking an average of 60 minutes to complete it. Participants were recruited through Prolific \cite{prolific_online} and compensated at an hourly rate of \$25. Before conducting the study with a larger group of participants, we piloted it to test the application's functionality and refined the vocabulary used in the study questionnaires.

\begin{table}[h]
\caption{Participant information for our mixed-methods user study.}
\label{tab:participants_study1}
 \scalebox{.85}{
\begin{tabular}{ll}
\toprule
& \textbf{Participant Groups}                                                 
\\ \midrule
\multirow{4}{*}{\begin{tabular}[c]{@{}l@{}}Age \\ Groups\end{tabular}} 
    & \begin{tabular}[c]{@{}l@{}}Below 25 years: 8 \end{tabular} \\ 
    & \begin{tabular}[c]{@{}l@{}}(25-35) years: 19 \end{tabular} \\ 
    & \begin{tabular}[c]{@{}l@{}}(36-45) years: 7 \end{tabular} \\ 
    & \begin{tabular}[c]{@{}l@{}}(46-55) years: 1 \end{tabular}                                                                                      
\\ \midrule
Gender & \begin{tabular}[c]{@{}l@{}}
Male: 18 \\
Female: 17 \\
\end{tabular}

\\ \midrule
\multirow{2}{*}{\begin{tabular}[c]{@{}l@{}}Highest \\ Education Level\end{tabular}} 
    & \begin{tabular}[c]{@{}l@{}}Bachelor: 21 \end{tabular} \\ 
    & \begin{tabular}[c]{@{}l@{}}Master: 9 \\ Doctorate: 5\end{tabular}

\\ \midrule
\multirow{2}{*}{\begin{tabular}[c]{@{}l@{}}Years of Experience with \\ Type 2 Diabetes Patients\end{tabular}} 
    & \begin{tabular}[c]{@{}l@{}}1-3 years: 15 \end{tabular} \\ 
    & \begin{tabular}[c]{@{}l@{}}3-5 years: 11 \\ 5-10 years: 5 \\ >10 years: 4\end{tabular} \\ 

\bottomrule
\end{tabular}}
\end{table}

\subsection{Participants}
The study involved 35 healthcare experts who were currently working in diverse roles, such as doctors, surgeons, nurses, medical researchers, paramedics, and medical assistants. They were recruited through Prolific. \Cref{tab:participants_study1} presents the demographic information of our study participants. To ensure that participants had vetted domain knowledge of type 2 diabetes, we implemented a custom screening questionnaire with three key inclusion criteria: (1) completion of the screening questionnaire with relevant responses to verify domain expertise, (2) at least six months of experience in treating and caring for diabetic patients, and (3) general awareness of the predictor variables used in the dataset and their impact on diabetes risk.  Moreover, the pre-screening questions enabled us to validate the healthcare professionals' backgrounds, ensuring they had obtained relevant certifications, licenses or training from accredited institutions for treating diabetic patients. These questions also assessed their understanding of type 2 diabetes pathophysiology, diagnostic criteria, treatment protocols, complication management, and lifestyle modification guidance. This questionnaire used to vet the domain expertise of the recruited participants was validated by healthcare experts from \anon{Faculty of Health Sciences, University of Maribor in Slovenia} and is available in the supplementary material. Before analysing the collected responses, we formulated two exclusion criteria: (1) responses from participants who did not answer all the study questions, and (2) responses from participants who did not complete the study tasks will be excluded. Additionally, given the growing concern about the use of large language models (LLMs) to cheat in crowd-sourced user studies \cite{veselovsky2023prevalencepreventionlargelanguage}, we planned to exclude responses that were entirely LLM generated as a post-study exclusion criteria. The participant responses were validated using Undetectable AI \cite{UndetectableAI} and GPTZero \cite{GPTZero} to identify any responses generated by LLM tools. Our validation process revealed no responses that were entirely generated by widely accessible LLMs. However, as indicated by the validation tools, only two participants appeared to have used an LLM for minor edits to some of their qualitative responses, with the key statements remaining their own. Moreover, the system log data indicated that they had completed the given study tasks, ensuring all other study protocols were followed. Consequently, no responses were excluded based on these criteria for our final study results.

\subsection{Evaluation Measures}
This section outlines the various measures collected in our user study to answer the research questions. The entire set of study questionnaires is available in the supplementary material.

\emph{Perceived trust}: To explore the impact of the debiasing process on trust in AI, we measured perceived trust using a multi-dimensional trust questionnaire adapted from Jian et al. \cite{Jian2020_trust_scale}. Multi-dimensional perceived trust refers to a critical user perspective for informed decision-making that incorporates various factors, including fidelity, reliability, and familiarity with the system or process. This measure was recorded on a 7-point Likert scale.

\emph{Metrics for evaluating participant performance in representation debiasing}: To evaluate the performance of participants in mitigating the representation bias within our system, we captured multiple objective measures through system logs. To assess the impact of the debiasing process on the AI system, we recorded the prediction model's accuracy, the overall quality score of the training dataset, and the estimated scores for various data issues. Following the methods proposed by Bhattacharya et al. \cite{bhattacharya2024exmos}, the dataset quality assessment did not account for the amount of representation bias. Instead, it focused on other data issues, such as outliers, redundant data, and correlated features, all of which were weighted equally in calculating the overall data quality score. The amount of representation bias was separately recorded through the overall representation rate (RR) and coverage Rate (CR) scores, as well as the RR and CR scores for each predictor variable after each retraining of the prediction model.  Furthermore, to investigate changes in the augmentation process performed by different participants, we captured the number of augmentations executed and the specific configurations used in the augmentation controller to generate new data.

\emph{Objective understanding (OU) of representation bias}: We also aimed to investigate whether participants' understanding of representation bias affected their ability to mitigate its impact. To assess their understanding, we measured objective mental model scores (i.e., objective understanding) of representation bias, following methods similar to those used by previous researchers \cite{kulesza_principles_2015, kulesza_explanatory_2010, bhattacharya2024exmos, Cheng2019, bove_contextualization_2022}. This measure was recorded on a scale of 0 to 5, where 0 represents the lowest OU score and 5 indicates the highest.

\emph{Subjective understanding of representation bias}: Additionally, we recorded participants' subjective understanding of representation bias using a perceived understandability questionnaire adapted from Hoffman et al.~\cite{hoffman2019metrics}. This measure was recorded on a 7-point Likert scale.

\emph{Perceived task load}: To assess whether the debiasing process was overwhelming for participants, we evaluated their perceived task load using the NASA-TLX questionnaire \cite{HART1988139, kulesza_principles_2015, kulesza_explanatory_2010}. The NASA-TLX includes six evaluation areas: mental demand, physical demand, time demand, performance, effort, and frustration. We followed the method proposed by Kulesza et al. \cite{kulesza_explanatory_2010}, where each area was rated on a scale from 0 to 20.

\emph{Qualitative questionnaire}: In addition to the previous measures, the participants were given open-ended qualitative questions regarding the UI components and their contribution to calibrating user understanding of representation bias, trust in the AI, and motivation of healthcare experts to actively participate in the debiasing process. 
\begin{figure*}
\centering
\includegraphics[width=0.95\linewidth]{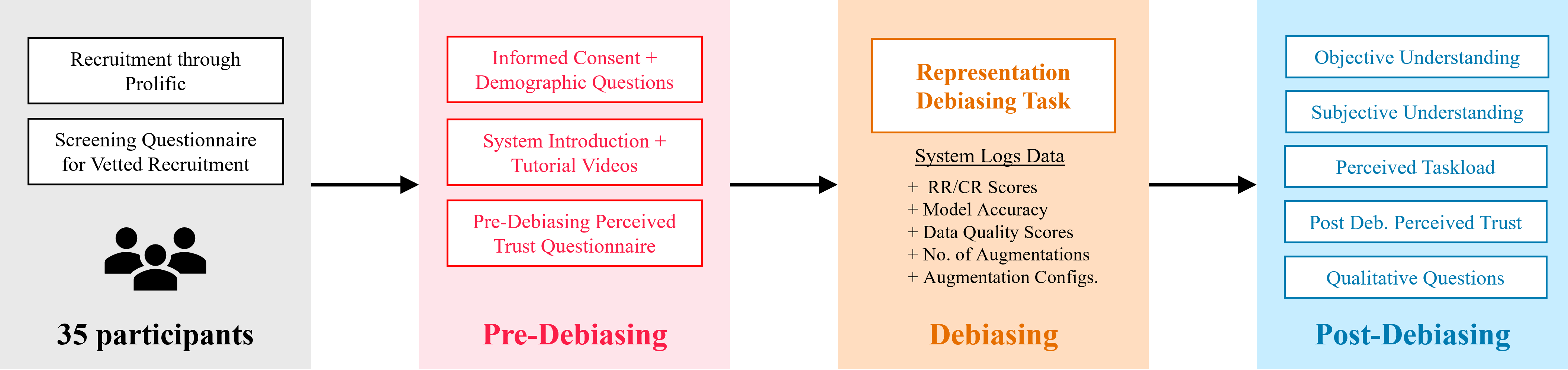}
\caption{Diagram illustrating the flow of our mixed-methods user study.}
\Description[User study flow]{Diagram illustrating the flow of our mixed-methods user study. The study included 35 healthcare experts who were recruited through Prolific through a screening questionnaire for vetted recruitment. The diagram also summarises the study flow and evaluation measures elaborated in \Cref{sec_user_study}.}
\label{fig:user_study_flow}
\end{figure*}

\subsection{Study Procedure}

Participants were first given detailed information about the study's objectives and their roles, responsibilities, and rights. They were required to provide informed consent in accordance with our ethical guidelines before proceeding. Following the informed consent, participants were asked to provide demographic information, summarised in Table 1. The overall study flow is illustrated in \Cref{fig:user_study_flow}.

Participants were then introduced to our debiasing application through detailed tutorial videos. These videos outlined the usage scenario, elaborately described the concept of representation bias and its general impact on prediction models, explained the role of the prediction model, and thoroughly demonstrated the functionality of each UI component described in \Cref{subsec_UI_components} for mitigating representation bias.

Next, they answered a pre-task questionnaire measuring their perceived trust in the AI system. This was done to establish a baseline, allowing us to measure how the debiasing process influenced their trust in the overall AI system.

Participants then proceeded to the main task of the study, which involved completing a debiasing task. In this task, at first, they were asked to explore the predictor variables to identify any signs of representation bias. Following this, they were asked to use the augmentation controller to generate a synthetic dataset aimed at addressing the identified biases. After generating the synthetic data, participants were required to validate and refine it before retraining the prediction model by merging the synthetic data with the original training set. This task had a maximum time limit of 30 minutes, during which participants could generate and retrain the model multiple times. Their primary objective was to reduce the amount of representation bias by maximising the representation rate (RR) and coverage rate (CR) scores without compromising the model accuracy and data quality.

After completing the primary task, participants were asked to fill out several post-task questionnaires centred on objective and subjective understanding of representation bias, perceived task load, trust in the AI system and several qualitative questions to justify their responses.

\subsection{Data Analysis}
To address our research questions, we analysed the collected data at multiple levels. First, we conducted an overall analysis considering all participants. Next, we wanted to investigate if different types of healthcare experts exhibited variations in the collected measures during their involvement in the debiasing process. To achieve this, we analysed the data by categorising participants into the following personas based on their healthcare roles: 
\begin{enumerate}[start=1,label={\textbf{ (\arabic*)}}, left=0.1cm, topsep=0pt, itemsep=0pt]
    \item \textit{Physicians} (including registered doctors, primary care physicians, general practitioners, and surgeons)
    \item \textit{Caregivers} (including registered nurses, home-care nurses, and emergency-care nurses)
    \item \textit{Other Healthcare Workers} (including paramedics, medical assistants, and medical researchers)
\end{enumerate}

\noindent Our participant pool consisted of 12 \textit{Physicians}, 12 \textit{Caregivers} and 11 \textit{Other Healthcare Workers}, categorised based on their current job roles. 

Additionally, we sought to determine whether varying levels of objective understanding of representation bias influenced the collected measures. To explore this, we grouped participants into the following three categories based on their objective understanding of representation bias:
\begin{enumerate}[start=1,label={\textbf{ (\arabic*)}}, left=0.1cm, topsep=0pt, itemsep=0pt]
    \item \textit{High Understanding} (OU Scores of 4 and 5) 
    \item \textit{Medium Understanding} (OU Scores of 2 and 3)
    \item \textit{Low Understanding} (OU Score less than 2) 
\end{enumerate}

\noindent Based on this grouping criteria, we had 19 participants with high objective understanding, 11 with medium objective understanding and 5 with low objective understanding.

Furthermore, to analyse the impact of domain experience on representation debiasing, we divided the participants into the following different experience levels based on their years of experience in administering type 2 diabetes patients:
\begin{enumerate}[start=1,label={\textbf{ (\arabic*)}}, left=0.1cm, topsep=0pt, itemsep=0pt]
    \item \textit{Higher Experience} (More than 5-years of experience) 
    \item \textit{Intermediate Experience} (3-5 years of experience)
    \item \textit{Lower Experience} (Less than 3 years of experience) 
\end{enumerate}
\noindent Considering the different experience levels, we had 9 participants with a higher experience level, 11 with an intermediate experience level and 15 with a lower experience level.

We conducted a series of statistical tests to analyse our data. Initially, we applied the Shapiro-Wilk normality test \cite{mccrum-gardner_which_2008} to determine the appropriate hypothesis testing methods, whether parametric or non-parametric. As the individual group data did not follow a normal distribution (confirmed by the Shapiro-Wilk test), we used the Kruskal-Wallis test \cite{mccrum-gardner_which_2008}, a non-parametric test, to assess the statistical significance of variations between groups based on the specified criteria. For subsequent pairwise comparisons between groups, we utilised the Mann-Whitney U-test with Bonferroni correction \cite{mccrum-gardner_which_2008}. Additionally, we assessed the correlation between factors, such as objective understanding of representation bias or change in perceived trust and RR/CR scores, by calculating Spearman's correlation coefficient \cite{mccrum-gardner_which_2008}. We applied a paired t-test \cite{mccrum-gardner_which_2008}, a parametric test, to determine the significance of changes in some of the evaluation measures (like perceived trust and accuracy) before and after the debiasing process when the data was found to be normally distributed using Shapiro-Wilk test.

Additionally, we investigated how a naive approach, not leveraging domain expertise, would affect the prediction model accuracy. To do so, we developed an automated approach for the data generation algorithm. This naive approach included a grid search tuning mechanism \cite{liashchynskyi2019gridsearchrandomsearch} aimed at maximising the RR and CR scores. We then compared the performance of this naive automated approach to the default model and evaluated if the participants were able to improve the model performance beyond these scores when involved in the debiasing process. 

Furthermore, we applied Braun and Clarke's thematic analysis method \cite{BraunClarkTA} for analysing the qualitative responses to generate themes that helped us understand the key patterns observed in the quantitative data in more depth.

\section{Results}

\subsection{How does involving domain experts during representation debiasing affect the prediction model's performance? (RQ1)}\label{sec_6_1_RQ1}

After the debiasing process, 24 out of 35 participants increased the prediction model's accuracy by an average of 2.1\%. The remaining participants maintained the same accuracy as the default value. Crucially, none of the participants reduced the model's default accuracy. After the debiasing process, to observe if the change in the model accuracy from the default model score was statistically significant, we used a paired t-test \cite{mccrum-gardner_which_2008} as the data was found to be normally distributed using the Shapiro-Wilk test. The results showed a statistically significant increase in model accuracy ($t = 7.35, p < .001$). 

\begin{figure*}
\centering
\includegraphics[width=0.8\linewidth]{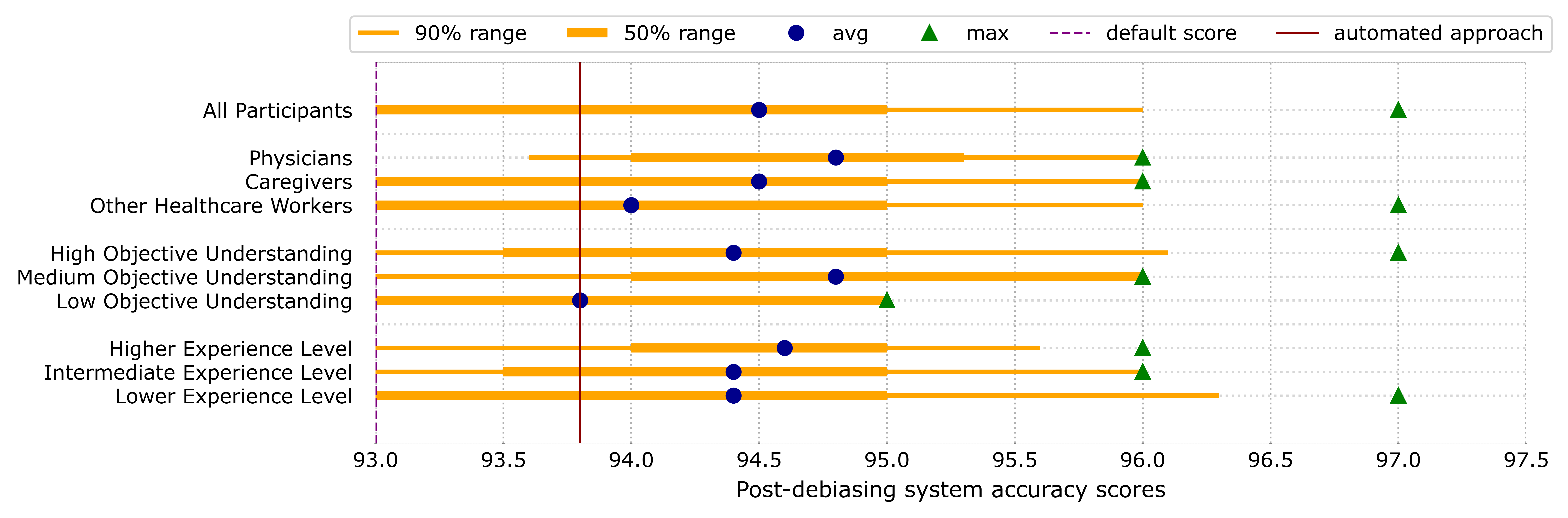}
\caption{Plot showing post-debiasing accuracy scores for all participants and for the different user groups. This plot presents the participant scores compared to the default model score and the naive automated approach scores.}
\Description[Post-debiasing accuracy scores]{Plot showing post-debiasing accuracy scores for all participants and for the different user groups. Overall, a majority of the participants were able to improve the accuracy of the prediction model. Amongst the different user groups, physicians achieved better post-debiasing accuracy compared to caregivers or other healthcare workers. However, this difference was not statistically significant. Similarly, participants with higher experience levels achieved higher accuracy than the other groups, but this difference was insignificant. No clear trend was observed based on different objective understanding levels. }
\label{fig:acc_score_changes}
\end{figure*}

The naive automated approach for applying the data augmentation algorithm to maximise the representation rate (RR) and coverage rate (CR) scores resulted in an accuracy of 93.8\%. This score was 0.86\% higher than the default model score. However, the average post-debasing accuracy by participants was slightly higher (by 1.3\%) than the accuracy obtained by the naive automated approach and was significant ($t=3.3, p = .001$). \Cref{fig:acc_score_changes} shows the participant accuracy scores compared to the default model score and the score obtained by the naive automated approach. 


\begin{figure}[h]
\centering
\includegraphics[width=1.0\linewidth]{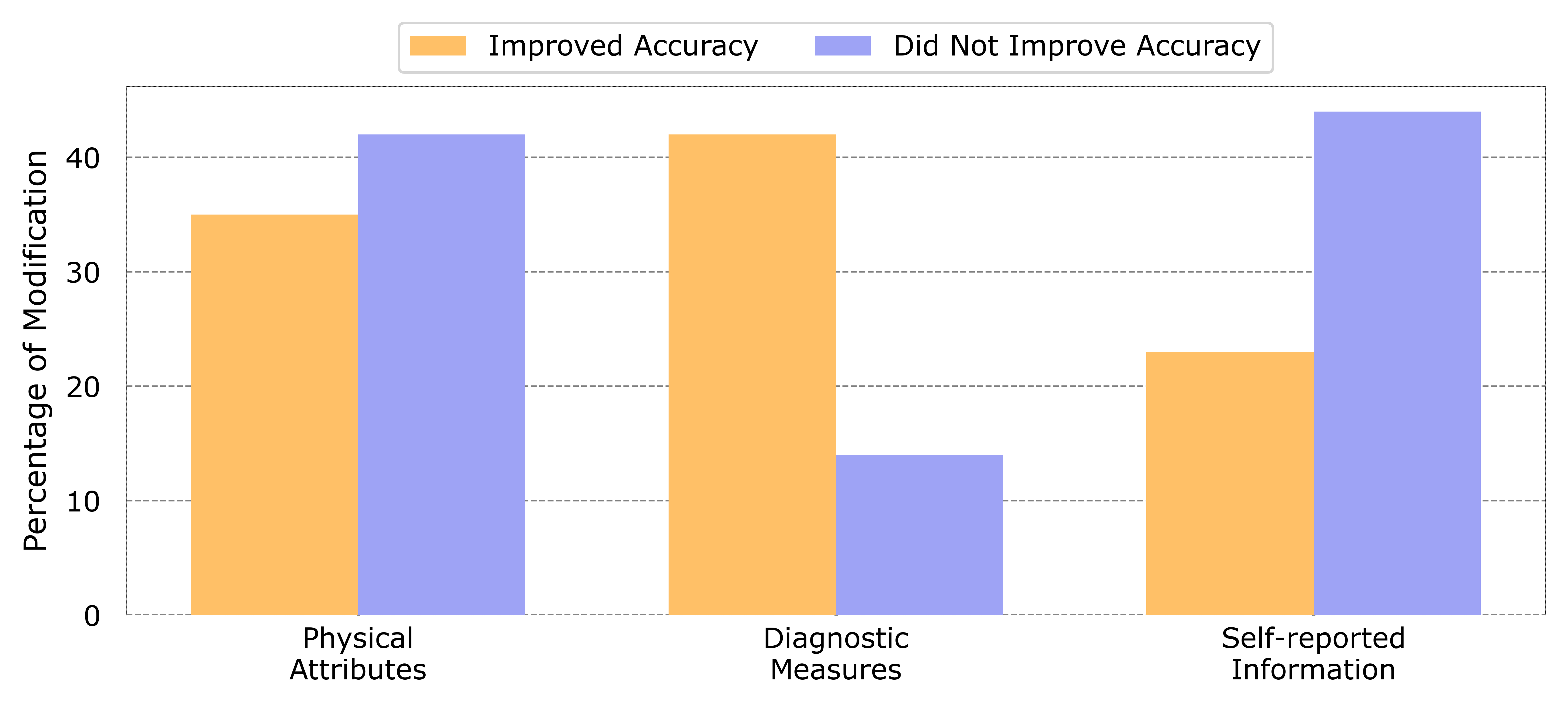}
\caption{Plot illustrating the average percentage of modifications made to the generated data across various predictor variable types. The comparison is between participants who successfully improved model accuracy after debiasing and those who could not improve the default accuracy level.}
\Description[Comparing different predictor variables edited for those who improved the accuracy vs those who could not improve the accuracy]{Plot illustrating the average percentage of modifications made to the generated data across various predictor variable types. The comparison is between participants who successfully improved model accuracy after debiasing and those who could not improve the default accuracy level.}
\label{fig:improved_vs_not_improved}
\end{figure}

We then analysed the system log data to understand why some participants were able to increase model accuracy while others were not. Interestingly, we found that participants who improved model accuracy made significantly more edits to diagnostic measures in the generated data (over 28\% more) compared to those who did not improve post-debiasing accuracy. This suggests that the data generation algorithm made more errors in generating values for diagnostic measures, and participants who identified and corrected these errors were able to increase the overall model accuracy. \Cref{fig:improved_vs_not_improved} shows a plot comparing the average percentage of modifications made to different types of predictor variables between participants who improved model accuracy and those who did not. Moreover, we did not observe any significant differences in the number of edits made between the two groups ($U=11, p=.841$), indicating that both participants who improved accuracy and those who did not improve, made a similar number of modification attempts.

\begin{figure}[h]
\centering
\includegraphics[width=1.0\linewidth]{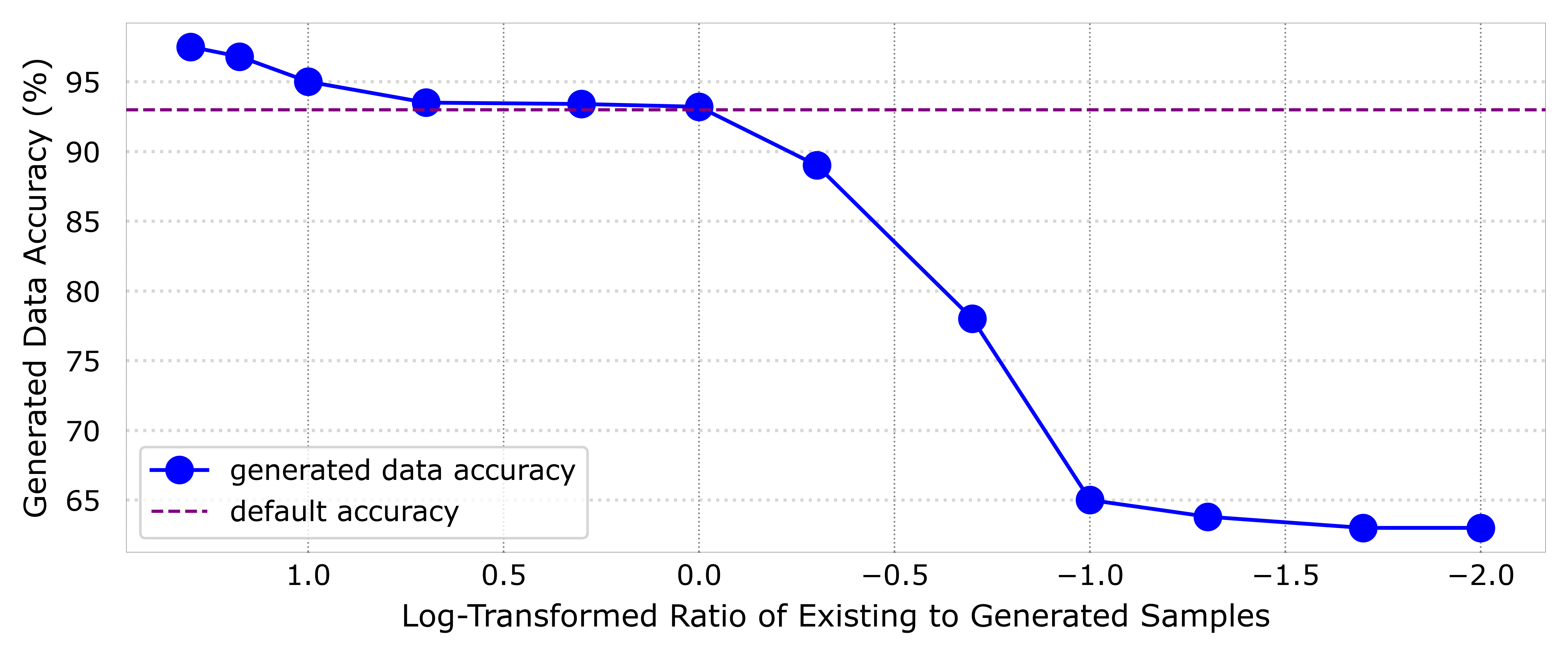}
\caption{Plot showing the accuracy levels of the default model on the generated data with respect to the log-transformed ratio of the number of existing samples to that of the number of samples to be generated.}
\Description[Accuracy Variation with Generated Data]{Plot showing the accuracy levels of the default model on the generated data with respect to the log-transformed ratio of the number of existing samples to that of the number of samples to be generated. This plot indicates that for the CTGAN algorithm to generate high-quality samples, the observation set size should be much higher than the size of the generated samples. Otherwise, if a high number of samples is generated from a small sample size of existing data, the generated data is expected to be more problematic.}
\label{fig:acc_vs_ratio}
\end{figure}

Another interesting observation emerged when we analysed the accuracy of the unedited generated data using the default trained model. We found that if participants used the augmentation controller to generate a very high number of synthetic samples from a small number of existing samples, the accuracy levels of the generated set were much lower. By examining the ratio of existing samples selected for data augmentation to the number of samples requested to be generated, we observed a decreasing trend in the accuracy levels of the generated data when the ratio was less than 1. \Cref{fig:acc_vs_ratio} illustrates how the accuracy of the generated data dropped when the log-transformed ratio of existing to generated samples fell below zero (i.e., when the ratio of the existing to generated sample size was less than 1). This indicates that when participants selected data segments with very low representation and attempted to increase the sample counts by an excessively higher amount to increase the RR/CR scores using the augmentation controller, the data generation algorithm produced problematic data points, leading to poor model accuracy. Additionally, we found that such generated datasets contained many outliers and abnormal records, requiring more effort and participant supervision to correct them.

Additionally, we inspected the qualitative data to understand the importance of domain experts in validating the data generated by the augmentation process from their perspective. We found many participants appreciated the ability to validate and refine the generated data. One of them mentioned: ``\textit{When I first created [generated] the new data table [synthetic data samples], I found that the estimated accuracy and the quality of the new data shown in the generated data controller to be very low compared to the [default] system scores. I noticed this problem when I selected certain categories in the augmentation controller with very few samples. I had to manually change the new data as not all the measures were consistent with each other. But after editing them, the estimated scores improved. This is a very useful feature of the application.}'' This response further supports our previous findings on the importance of domain experts participating in representation debiasing, especially for validating synthetic data generated from extremely low-represented data segments and refining values for certain predictor variables that require deep domain knowledge (such as diagnostic patient measures).

However, we did not observe any significant difference in the post-debiasing accuracy across different personas ($H= 3.41, p = .18$), objective understanding levels ($H = 2.99, p = .22$), or experience levels ($H = 0.13, p = .935$). This result indicates that the change in post debiasing accuracy is consistent across different types of participants.  \Cref{fig:acc_score_changes} presents the post-debiasing accuracy scores for all participants and the different participant groups. Although we did observe a slight variation in the post-debiasing accuracy scores amongst the different participant groups, the difference was not statistically significant. However, the post-debiasing accuracy was strongly correlated with the representation rate (RR) ($r=0.35, p = .041$) and coverage rate (CR) scores ($r= 0.35, p = .038$), indicating that participants who improved the model performance also achieved a significant reduction in representation bias.

\colorlet{framecolor}{main}
\colorlet{shadecolor}{sub}
\setlength\FrameRule{0pt}
\begin{frshaded*}
\noindent\textbf{Key-takeaways}: Involving domain experts in the debiasing process does not compromise the model performance. Their involvement is particularly important for validating synthetic data generated from extremely low-represented data segments and refining values for certain predictor variables that require deep domain knowledge.
\end{frshaded*}

\subsection{How can domain experts contribute to reducing representation bias in AI/ML systems? (RQ2)}\label{sec_6_2_RQ2}

\begin{figure*}
\centering
\includegraphics[width=1.0\linewidth]{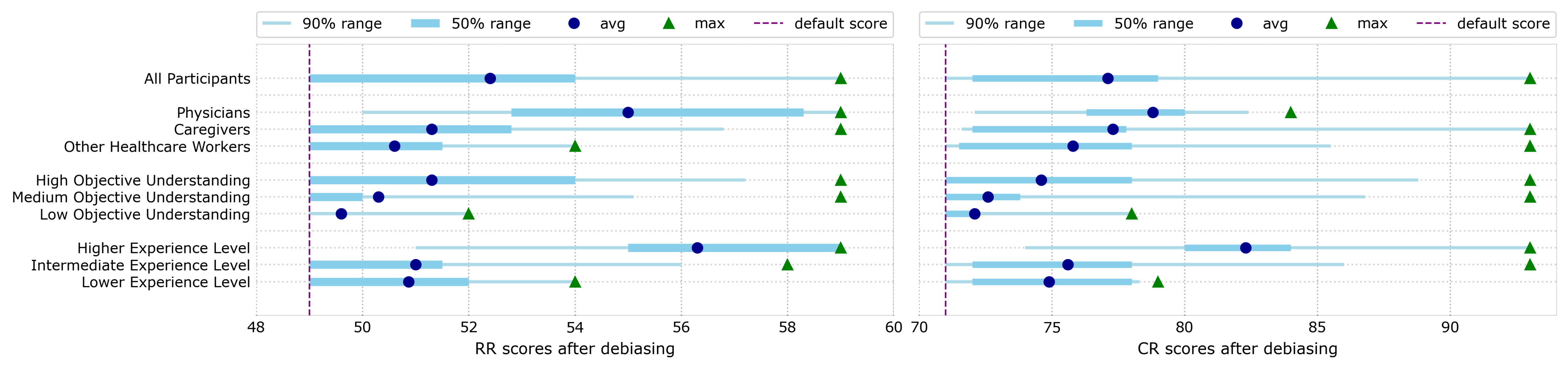}
\caption{Representation rate (RR) and coverage rate (CR) scores for user study participants after the debiasing task.}
\Description[Representation rate (RR) and coverage rate (CR) scores for user study participants.]{Representation rate (RR) and coverage rate (CR) scores for user study participants after the debiasing task. The average increase in RR scores was approximately 4\%, and the increase in CR scores was approximately 6\%. The graph shows a clear trend between the different user groups based on persona, objective understanding and experience levels for both RR and CR scores.}
\label{fig:debiasing_scores}
\end{figure*}

Based on our study results, 31 out of 35 participants improved the representation rate (RR) and the coverage rate (CR) scores, with an average increase of approximately 4\% in RR and 6\% in CR scores. Importantly, none of them worsened these scores following the debiasing process, although 4 participants were unable to improve the default scores. Upon analysing these results based on the different personas, using a Kruskal-Wallis test, we observed a significant difference in the RR scores ($H = 10.62, p = .004$), as shown in \Cref{tab:stats_summary_rq1}. Further analysis with the Mann-Whitney U-test adjusted with Bonferroni corrections (as summarised in \Cref{tab:stats_summary_rq1_table2}) revealed that \textit{Physicians} were significantly better than \textit{Caregivers} ($ U=114.0, p=.015$) and \textit{Other Healthcare Workers} in improving the RR scores ($U=115.5, p=.002$). However, the difference in CR scores was not statistically significant ($H = 3.48, p = .17$), even though \Cref{fig:debiasing_scores} shows a clear trend in the mean CR scores, with \textit{Physicians} achieving a higher mean CR score compared to the other groups. Moreover, no significant differences were observed between the groups in the number of data augmentations performed ($H = 1.69, p = .43$). This suggests that the number of data augmentation attempts was relatively similar across all groups. 

\begin{table*}
\centering  
\caption{Summary of statistical significance assessments using Kruskal-Wallis test for the different participant groups for RR/CR scores and number of augmentations.}
\label{tab:stats_summary_rq1}
\scalebox{.7}{
\begin{tabular}{@{}ccc@{}}
\toprule
\textbf{Participant Groups}                                                                                                                               & \textbf{Measures}                & \textbf{Kruskal-Wallis Test Scores}  \\ \midrule
\multirow{4}{*}{\begin{tabular}[c]{@{}c@{}}\underline{\textit{Persona}}: \\ \textsc{Physicians (n=12)}, \\ \textsc{Caregivers (n=12)}, \\ \textsc{Other Healthcare Workers (n=11)}\end{tabular}} 

& & \\
& \cellcolor{lightgray}RR Scores               &   \cellcolor{lightgray}$H = 10.62, p = .004$         \\
                                                                                                                                                 & CR Scores               & $H = 3.48, p = .17$           \\
                                                                                                                                                 & Number of Augmentations & $H = 1.69, p = .43$  \\         \\ \cmidrule(l){1-3} 
\multirow{4}{*}{\begin{tabular}[c]{@{}c@{}}\underline{\textit{Objective Understanding}}: \\ \textsc{High (n=19)}, \\ \textsc{Medium (n=11)}, \\ \textsc{Low (n=5)}\end{tabular}}                 
& & \\
& \cellcolor{lightgray}RR Scores               & \cellcolor{lightgray}$H = 14.72, p < .001$ \\
                                                                                                                                                 & \cellcolor{lightgray}CR Scores               & \cellcolor{lightgray}$H = 11.09, p = .003$         \\
                                                                                                                                                 & Number of Augmentations &         $H = 3.39, p = .18$               \\     \\ \cmidrule(l){1-3} 
\multirow{4}{*}{\begin{tabular}[c]{@{}c@{}}\underline{\textit{Experience Levels}}: \\ \textsc{Higher (n=9)}, \\ \textsc{Intermediate (n=11)}, \\ \textsc{Lower (n=15)}\end{tabular}}                 
& & \\
& \cellcolor{lightgray}RR Scores               & \cellcolor{lightgray}$H = 12.27, p = .002$         \\
                                                                                                                                                 & \cellcolor{lightgray}CR Scores               & \cellcolor{lightgray}$H = 9.99, p = .006$          \\
                                                                                                                                                 & Number of Augmentations &    
                                                                 $H = 1.38, p = .50$
                                                                                \\
                                                                                                                                                 
\\
\bottomrule
\end{tabular}}
\end{table*}

\begin{table*}
\centering  
\caption{Summary of statistical significance assessments using Mann-Whitney U-test for the different participant groups. After adjusting the \textit{p-value} with Bonferroni correction, $p < .0167$ is considered to be significant.}
\label{tab:stats_summary_rq1_table2}
\scalebox{.7}{
\begin{tabular}{@{}llll@{}}
\toprule
\textbf{Measures}                & \multicolumn{3}{c}{\textbf{Mann-Whitney   U-test scores}}                                                   \\ \midrule
                        & \textit{Physician-Caregiver}             & \textit{Physician-OHW }           & \textit{Caregiver-OHW}                  \\ \cmidrule(l){2-4} 
RR Scores               & \cellcolor{lightgray}$U=114.0, p=.015$                 & \cellcolor{lightgray}$U=115.5, p=.002$          & $U=68.5, p=.895$                 \\
CR Scores               & $U=91.5, p=.268$                 & $U=95.5, p=.070$          & $U=77.0, p=.503$                \\
Number of Augmentations & $U=85.0, p=.101$                 & $U=66.0, p=.231$          & $U=56.0, p=.808$                \\ \midrule
                        & \textit{High OU - Medium OU}             & \textit{High OU - Low OU}         & \textit{Medium OU - Low OU}             \\ \cmidrule(l){2-4} 
RR Scores               & \cellcolor{lightgray}$U=3572.5, p=.0145$                 & \cellcolor{lightgray}$U=2330.5, p<.001$          & $U=835.0, p=.165$                \\
CR Scores               & $U=3415.0, p=.065$                 & \cellcolor{lightgray}$U=2272.5, p=.002$          & $U=849.0, p=.122$                \\
Number of Augmentations & $U=72.5, p=.679$                 & $U=53.5, p=.146$          & $U=30.5, p=.063$                \\ \midrule
                        & \textit{Higher Exp. - Intermediate Exp.} & \textit{Higher Exp. - Lower Exp.} & \textit{Intermediate Exp. - Lower Exp.} \\ \cmidrule(l){2-4} 
RR Scores               & \cellcolor{lightgray}$U=86.5, p=.004$                 & \cellcolor{lightgray} $U=121.5, p=.001$          & $U=77.5, p=.89$                \\
CR Scores               & \cellcolor{lightgray}$U=82.0, p=.014$                 & \cellcolor{lightgray}$U=117.5, p=.002$          & $U=80.0, p=.91$                \\
Number of Augmentations & $U=22.5, p=.34$                 & $U=45.5, p=.36$          & $U=49.5, p=.51$                \\ \bottomrule
\end{tabular}
}
\end{table*}

\begin{figure*}
\centering
\includegraphics[width=1.0\linewidth]{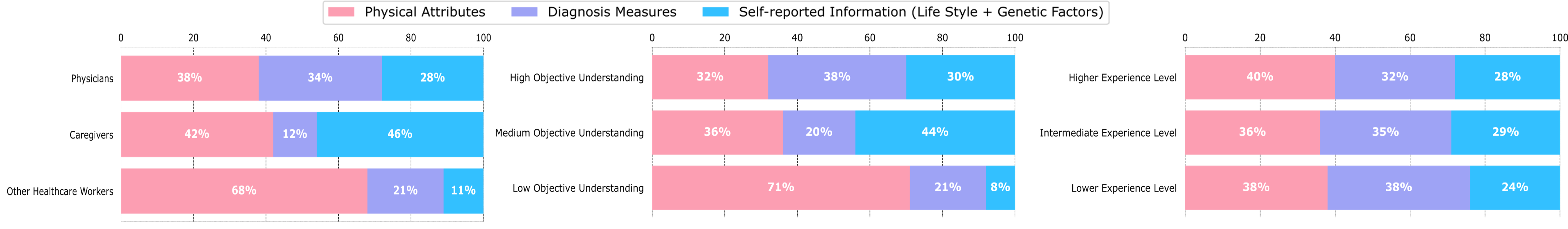}
\caption{Plot comparing the proportions of different types of predictor variables selected during the augmentation process, highlighting differences across user groups. }
\Description[Plot showing the proportion of different types of variables selected during the augmentation process for each user group.]{Plot showing the proportion of different types of variables selected during the augmentation process for each user group. Considering the different personas, only physicians gave high importance to diagnosis measures compared to caregivers and other healthcare workers. Similarly, trends were observed for participants with a higher objective understanding. However, no significant trend was observed for the participants with different experience levels. }
\label{fig:variable_type_altered}
\end{figure*}

To better understand why \textit{Physicians} excelled at reducing overall representation bias, we analysed the differences in the types of predictor variables selected during the data augmentation process. We observed that \textit{Physicians} gave almost equal importance to all types of predictor variables during the augmentation process. In contrast, \textit{Caregivers} paid less attention to diagnosis measures, while \textit{other healthcare workers} placed less emphasis on lifestyle variables during data augmentation. As illustrated in \Cref{fig:variable_type_altered}, this result suggests that \textit{Physicians}' robust domain knowledge of the significance of diverse health variables led them to emphasise a broader range of predictor variables for the debiasing process. Particularly, they paid much more attention to diagnosis measures than the other two groups. Consequently, they performed better during the debiasing task.

Inspecting the results from the perspective of objective understanding of representation bias provides preliminary evidence that domain experts with a higher objective understanding of representation bias can be more effective at mitigating its impacts. However, since objective understanding was measured after participants interacted with the system during the debiasing task, this does not imply a direct causal relationship. We found a significant correlation between objective understanding and both RR ($r=0.54, p<.001$) and CR scores ($r=0.44, p=.007$). We also found a significant difference in the RR scores ($H = 14.72, p < .001$) and CR scores ($H = 11.09, p = .003$) between the groups having different levels of objective understanding. Particularly, participants with a higher objective understanding level had a higher RR ($ U=2330.5, p<.001$) and CR scores ($ U=2272.5, p=.002$) than those with a lower objective understanding. Consistent with the findings across different persona groups, no significant differences in the number of augmentation attempts were observed among groups with varying levels of objective understanding. Additionally, the \textit{High Objective Understanding} group exhibited similar patterns in the types of predictor variables controlled during the augmentation process as those observed in the Physicians group, as shown in \Cref{fig:variable_type_altered}. 

Further analysis based on the domain experience levels of the participants revealed significant differences in both RR ($H = 12.27, p = .002$) and CR scores ($H = 9.99, p = .006$) using the Kruskal-Wallis test between the three groups. From the Mann-Whitney U-test results with Bonferroni corrections, we found that participants with a higher experience level had better RR and CR scores than those with intermediate [RR: ($U=86.5, p=.004$), CR: ($U=82.0, p=.014$)] and lower experience levels [RR: ($U=121.5, p=.001$), CR: ($U=117.5, p=.002$)]. However, no notable differences were observed either in the number of augmentations performed ($H=1.38, p=.50$) or the types of predictor variables controlled during the task.

While these results suggest that domain experts can contribute to reducing the representation bias in ML systems, surprisingly, we found a significant decrease in the overall data quality scores after the debiasing process ($t=-2.12, p=.029$). Upon further investigation, this change was primarily driven by changes observed in the data outlier and feature correlation scores. Interestingly, the amount of data outliers was significantly lower after the debiasing process ($t=-5.10, p<.001$). This result suggested that the debiased data has fewer outliers and was better for the prediction model. However, there was a significant increase in feature correlation scores post-debiasing ($t=1.8, p=.036$), which largely contributed to the decrease in the overall quality scores. 

Analysis of the qualitative responses revealed the following themes that helped us understand the underlying reasons for the observed increase in correlated features post-debiasing:

(1) \emph{\textbf{Difference in natural correlations in clinical factors than observed correlation in training data}}: The qualitative participant responses revealed that certain predictor variables could be inherently associated with other variables in larger sample sizes. However, these natural correlations might not be accurately reflected in the training data. As mentioned by one of the participants: ``\textit{Patients in real life with diabetes mostly share similar clinical findings such as BMI above 30, systolic blood pressure raised. There will be some bias due to the contributing factors of the disease.}'' This response highlights that certain clinical factors (such as BMI and systolic blood pressure) are inherently correlated due to the nature of the disease (diabetes) and its contributing factors. Consequently, real-world associations and correlations among predictor variables may differ from those observed in training datasets.

(2) \emph{\textbf{Debiasing process can lead to a more accurate representation of real-world feature correlations than what is observed in both training or generated data}}: Participants mentioned that validating and refining the generated data can lead to more accurate representations of real-world observations. One of them mentioned, ``\textit{I have validated some of the samples with weird values like a very low BMI, yet very high SBP [systolic blood pressure]. Generally, patients with low BMI have low blood pressure and not higher.}'' Participants refined the generated data by considering clinical associations of health variables. The overall feature correlation likely increased after the debiasing process because it revealed and corrected inherent clinical associations and anomalies that were initially absent in the original training data. Consequently, these natural correlations became more apparent post-debiasing, leading to an increase in observed feature correlation.

\colorlet{framecolor}{main}
\colorlet{shadecolor}{sub}
\setlength\FrameRule{0pt}
\begin{frshaded*}
\noindent\textbf{Key-takeaways}: Results demonstrate that domain experts can effectively contribute to mitigate representation bias. Particularly, domain experts with robust knowledge of all predictor variables, a higher objective understanding of representation bias and a higher domain experience level can be more efficient in representation debiasing. Additionally, their involvement can lead to a more accurate representation of real-world feature correlations than observed in training and generated data.
\end{frshaded*}

\subsection{How does the involvement of domain experts in the representation debiasing process influence their trust in AI systems? How could the debiasing process be improved? (RQ3)}

\begin{figure}[h]
\centering
\includegraphics[width=1.0\linewidth]{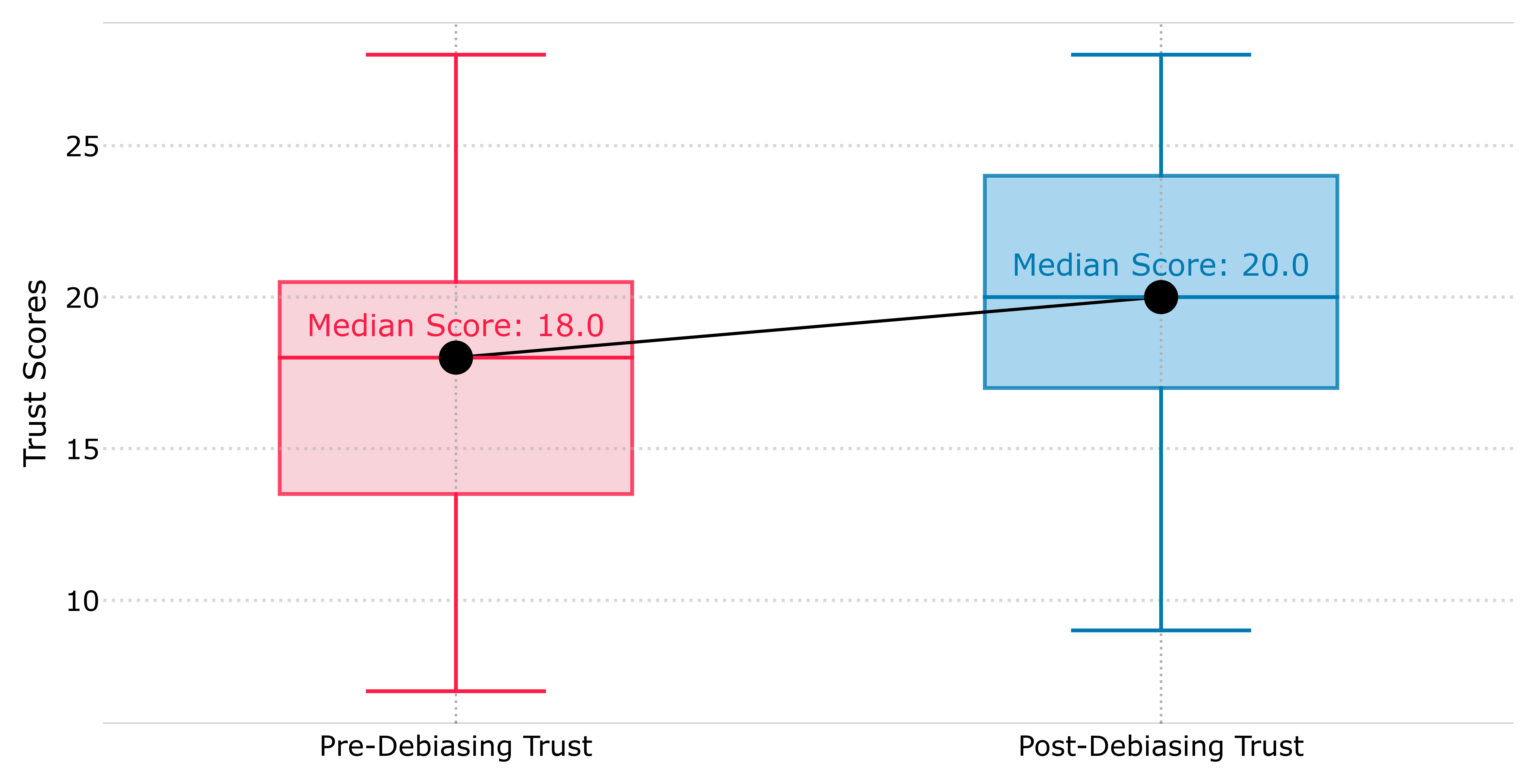}
\caption{Box plots showing an increase in the perceived trust scores after the debiasing process.}
\Description[Change in Perceived Trust scores]{Box plots showing an increase in the perceived trust scores after the debiasing process. The post-debiasing trust scores were higher by 7\% on average compared to the pre-debiasing scores.}
\label{fig:perceived_trust_scores}
\end{figure}

Using a paired t-test, we found that post-debiasing trust scores were significantly higher than the pre-debiasing trust scores ($t=-2.47, p=.009$). This increase in trust after the debiasing process was correlated with both subjective understanding  ($r = 0.58, p<.001$) and objective understanding of representation bias ($r=0.34, p=.004$). We also observed a strong correlation between perceived trust and RR scores ($r=0.4, p=.016$). However, the correlation between perceived trust and CR scores ($r=0.33, p=.053$) was insignificant despite observing an increasing trend.  These results suggest that as domain experts reduce the impact of representation bias, their trust in AI may increase. \Cref{fig:perceived_trust_scores} shows the increase in overall perceived trust scores after the debiasing process. The post-debiasing trust scores were, on average, approximately 7\% (or 2 points) higher than the pre-debiasing scores.

\begin{table*}
\centering  
\caption{Summary of statistical test scores for the change in perceived trust measure after the debiasing process for the different participant groups.}
\label{tab:stats_summary_rq2}
\scalebox{.8}{
\begin{tabular}{@{}ccc@{}}
\toprule
\textbf{Different Group Types}                                                                                                                               & \textbf{Participant Groups}                & \textbf{Paired t-test scores}  \\ \midrule
\multirow{3}{*}{\begin{tabular}[c]{@{}c@{}}\textit{Persona} \end{tabular}} 

& \cellcolor{lightgray}\textsc{Physicians (n=12)}               &   \cellcolor{lightgray}$t=-2.17, p = .026$         \\
                                                                                                                                                 &
                                                                                                                                                 \cellcolor{lightgray}
                                                                                                                                                 \textsc{Caregivers (n=12)}             & 
                                                                                                                                                 \cellcolor{lightgray}
                                                                                                                                                 $t=-2.08, p = .030$           \\
                                                                                                                                                 & \textsc{Other Healthcare Experts (n=11)}
                                                                                                                                                 & $t=0.68, p=.745$  \\  \midrule 
\multirow{3}{*}{\begin{tabular}[c]{@{}c@{}}\textit{Objective Understanding} \end{tabular}}                 
& \cellcolor{lightgray}\textsc{High (n=19)}               & \cellcolor{lightgray}$t=-2.66, p = .007$ \\
                                                                                                                                                 & \textsc{Medium (n=11)}               & $t=-0.27, p = .395$         \\
                                                                                                                                                 & \textsc{Low (n=5)} &         $t=-0.12, p = .45$     \\ \midrule
\multirow{3}{*}{\begin{tabular}[c]{@{}c@{}}\textit{Experience Levels} \end{tabular}}                 
& \cellcolor{lightgray}\textsc{Higher (n=9)}               & \cellcolor{lightgray}$t=-3.97, p = .002$         \\
                                                                                                                                                 & \textsc{Intermediate (n=11)}              & $t=-1.24, p = .121$          \\
                                                                                                                                                 & \textsc{Lower (n=15)} &    
                                                                 $t=2.04, p = .97$
                                                                   
\\
\bottomrule
\end{tabular}}
\end{table*}

Moreover, this increase in perceived trust was found to be significant for \textit{Physicians} ($t=-2.17, p=.026$) and \textit{Caregivers} ($t=-2.08, p=.03$), as shown in \Cref{tab:stats_summary_rq2}. Considering the different objective understanding levels, the change in perceived trust was significantly high only for the group with a higher objective understanding ($t=-2.66, p=.007$). Considering the different experience levels, we observed that the increase in perceived trust post-debiasing is significant for the group with the higher experience level ($t=-3.97, p=.002$).

Findings from the qualitative responses provided insights into the factors that contributed to higher perceived trust among participants following the debiasing process, which are summarised in the following themes:

(1) \emph{\textbf{Importance of data-centric explanations in understanding representation bias}}: From the qualitative responses, participants frequently emphasised the importance of data-centric explanations in the Data Explorer component for understanding representation bias. They noted that providing information about data distributions, comparing multiple predictor variables, and highlighting the impact of representation bias on prediction outcomes through these explanations made it easier to identify and mitigate representation bias: ``\textit{The system really helped me understand representation bias by showing me how different groups are represented in the data. It made it clear how well the model was performing for each group and if there were any imbalances.}'' As data-centric explanations increased their understanding of the impact of representation bias, participants' overall trust in the AI appeared to increase as a result.

(2) \emph{\textbf{Importance of giving domain experts control to refine generated data}}: As mentioned earlier, participants valued the system’s ability to let them inspect and refine generated data, which improved their understanding of representation bias and its impact on the model performance.  They recognised that blindly adding more samples does not effectively address representation bias. One participant noted, ``\textit{After re-training the AI with validated data, my understanding of representation bias has deepened. This process underscored that simply adding data is insufficient. It must accurately reflect real-world diversity to improve fairness, accuracy, and reliability.}'' Providing them with control to validate and refine generated data, helped them to build more trust in the AI system.

(3) \emph{\textbf{Transparency of the generated data quality is essential for building trust}}: Building on the previous theme, participants reported that the transparency of the generated data, through the display of estimated quality and prediction accuracy, helped them develop more confidence and trust during the debiasing process. One of the mentioned, ``\textit{Generating and validating new data makes me trust the AI system more because it shows the quality and the model accuracy on the new data. When the new data helps the model perform well for all groups, it means the system is more accurate and reliable. Overall, knowing that the system uses validated data to enhance its performance and fairness makes me feel more confident in its results.}''

(4) \emph{\textbf{Collaboration between Domain Expert and AI during the debiasing process is crucial for building their trust in the AI}}: Participants mentioned that AI and domain experts could complement each other during the debiasing process, as only AI algorithms might be insufficient for the debiasing process. One of them mentioned, ``\textit{An expert might be able to mitigate these biases based on an underlying knowledge of circumstances, but AI does not have access to this knowledge without programming. I would not trust an AI system basing its output on data with much bias.}'' Another participant mentioned, ``\textit{The representation bias in the system makes me a bit sceptical about fully relying on its help; it needs to be backed up with proper clinical reasoning by an experienced clinician. The two can complement each other.}''

\begin{figure}[h]
\centering
\includegraphics[width=1.0\linewidth]{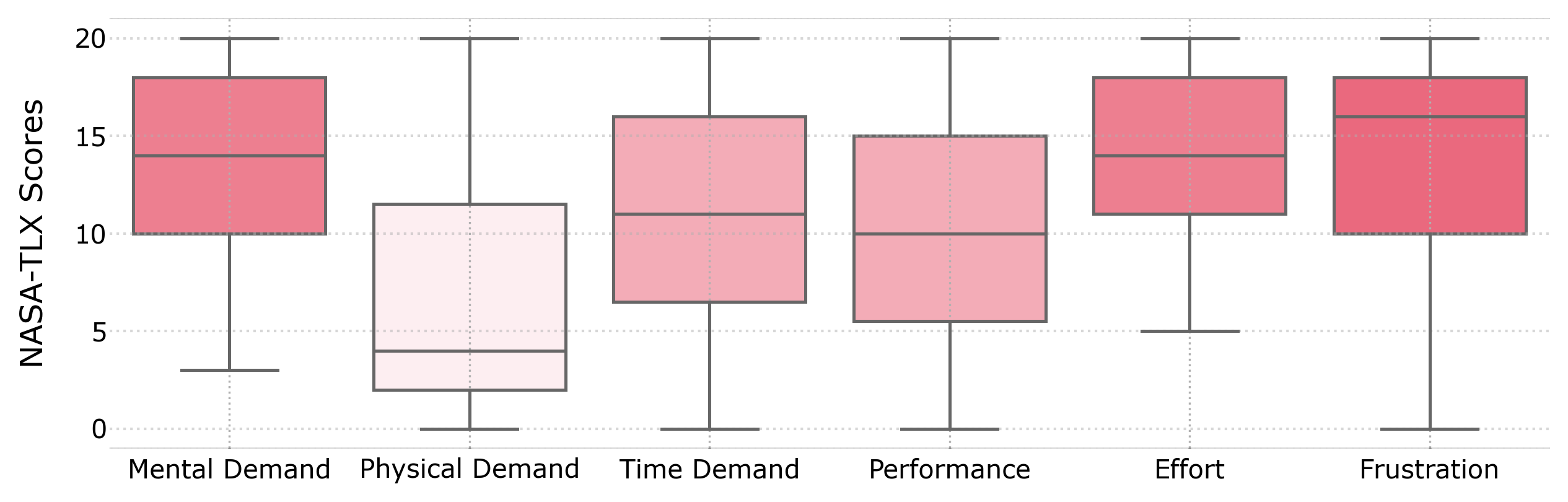}
\caption{NASA-TLX scores for perceived task-load assessment across multiple factors.}
\Description[NASA-TLX scores for perceived task-load assessment across multiple factors.]{NASA-TLX scores for perceived task-load assessment across multiple factors. From the NASA-TLX assessment, we can infer that the frustration levels, mental demand and effort levels of the users were elevated. }
\label{fig:nasa_tlx_scores}
\end{figure}

Trust is one aspect of the user experience, so we turned our attention to other measures that capture the acceptability of the debiasing process. The results of the NASA-TLX assessments, as shown in \Cref{fig:nasa_tlx_scores}  indicate that the mental demand, effort and frustration levels were slightly elevated, suggesting that there is scope to further improve the user experience of domain experts during the debiasing process. We analysed the qualitative responses to better understand potential improvements that can alleviate the elevated mental demand and frustration levels experienced by domain experts during representation debiasing. Consequently, the following themes generated using a thematic analysis of the qualitative response highlight potential improvements to enhance the overall user experience of domain experts during the debiasing process:

(1) \emph{\textbf{Need for a global overview of bias for specific patient groups and associated health variables}}: Our participants expressed a need for a global overview of representation bias for specific patient groups. The current UI only allowed users to explore bias in individual variables, making it challenging to switch between different variables to assess the impact of the bias on specific segments: ``\textit{the understanding of representation bias can be improved by adding better tools [UI components] that show how different groups are performing.}'' Additionally, they added that a chatbot or a query system could be easier to analyse the presence of bias in specific segments: ``\textit{A chatbot or a query system can be better instead of exploring all the variables. If I enter the specifications of the patient group, it should show me the amount of representation bias for that specific patient group.}'' Furthermore, some participants expressed a need for comparing multiple associate variables in the same view: ``\textit{What I am missing is a pairwise comparison. For example, BMI could be associated with factors like their family history of diabetes or drinking level. It is a bit difficult to compare each factor individually.}''

(2) \emph{\textbf{Ability to select variables or sub-groups where the impact of the bias is more important than other segments}}: Participants pointed out that representation bias does not affect all health variables or their sub-groups equally. They also indicated that achieving a completely balanced distribution for all variables and sub-groups is not practically feasible.  Thus, they expressed a need for selecting specific variables or segments they consider more important when computing the overall bias scores: ``\textit{It is not realistic to expect that all the sub-categories will be balanced and bias free. There should be an option to select categories or sub-categories that are expected to be well-balanced. Other factors can have only a few samples, but they need not well balanced. The only thing that should be ensured is that some data from all patient groups is available so that the AI can be used with all types of patients}.''

(3) \emph{\textbf{Additional AI assistance during refinement of generated data}}: To further assist domain experts in identifying problematic generated samples, participants suggested that AI assistance, such as a chatbot or automated algorithms highlighting potential issues in the generated data, would be very helpful. One of them stated, ``\textit{AI assisted detection of problematic generated data using some statistical algorithm would make it easier for me to validate and streamline the generated data.}'' Another one added, ``\textit{A summary of the misleading data values for the factors in the generated data can help in identifying problematic samples. A visual summary of such samples or even a chatbot to inspect the generated could be very handy for GPs.}''

\colorlet{framecolor}{main}
\colorlet{shadecolor}{sub}
\setlength\FrameRule{0pt}
\begin{frshaded*}
\noindent\textbf{Key-takeaways}: Involving domain experts in the debiasing process can increase their perceived trust in the AI, especially when they successfully reduce representation bias by thorough validation of the generated data. Qualitative insights from our study highlight several factors that contribute to building trust through such a debiasing process and certain improvements that can alleviate the elevated mental demand, effort and frustration levels experienced by domain experts during representation debiasing.
\end{frshaded*}

\section{Discussions}

\subsection{Limitations}
Before discussing the implications of our research, we would like to acknowledge the following known limitations of our work:
\begin{enumerate}[start=1,label={ (\arabic*)}, left=0.1cm]
    \item \textit{Lack of restrictions on generated data volume}:  Our debiasing application did not impose limits on the quantity of generated samples. Consequently, some participants produced over 50,000 samples from a relatively small training dataset. This excessive sample generation led to slow performance during the validation and re-training phases. Additionally, generating an extremely high amount of samples can also cause the application to crash due to memory overload issues. To address these issues, we recommend that future implementations of such systems establish strict upper limits on the volume of data that can be generated.
    \item \textit{Additional UI improvements for validating generated data}: In connection with the previous limitation, when users generated a large volume of data, the existing UI controls for filtering and identifying problematic samples proved insufficient as users found it difficult to identify these problematic samples using the given UI controls. Future implementations should incorporate enhanced UI components that enable global exploration and visualisation of the generated data, facilitating more effective validation.
    \item \textit{Potential improvements in the participant recruitment process}:  We relied on Prolific's background verification process to recruit verified healthcare experts\footnote{\url{https://www.prolific.com/specialist-participants}} and did not manually verify the credentials or professional identities of the participants. To enhance the reliability of participant recruitment (especially domain expert recruitment) from crowd-sourcing platforms like Prolific, we recommend future researchers to implement additional verification steps, such as cross-checking professional license numbers or identification cards issued by their affiliated institutions. Despite our efforts to recruit vetted domain experts from Prolific using a screening questionnaire and strict inclusion and exclusion criteria, we acknowledge the possibility of some participants relying on LLM-based tools to provide responses, potentially affecting the authenticity and unbiased nature of their perspectives. Additionally, participants should be explicitly instructed not to use any LLM tools during their participation in the study.
\end{enumerate}

\subsection{Implications for Research, Practice and Design}
\subsubsection{Benefits of Involving Domain Experts in Representation Debiasing}

Although the extent of user control in human-AI collaboration has been a subject of debate \cite{konig2022challenges, Westphal2023, Yang2020,  zha2023datacentric}, our study results provide clear evidence that domain experts can be integrated into the debiasing process of representation bias. In fact, their involvement is crucial in validating generated data, particularly when samples are generated from extremely underrepresented data segments or for predictor variables that require nuanced domain knowledge to understand and ensure the preservation of natural associations and relationships (as noted in our findings presented in \Cref{sec_6_1_RQ1} and \Cref{sec_6_2_RQ2}). Despite the popularity of generative AI algorithms, supervision by domain experts remains essential in high-stakes domains due to potential issues such as the generation of noisy or corrupt samples that do not accurately reflect real-world observations \cite{elor2022smotesmote, MUMUNI_DAC, balestriero2022effects, superb-ai-blog}. Moreover, as observed in our findings, domain experts can also expose hidden feature correlations and other data issues that automated detection algorithms might miss, as these issues may not always be statistically significant but can still impact the prediction model. In fact, their inputs can be valuable in confirming if the training dataset consists of potential bias or issues considering real-world observations.  

Consistent with prior works \cite{kulesza_principles_2015, bhattacharya2024exmos}, our findings also demonstrate that involving domain experts does not compromise the performance of AI models. Instead, they can improve accuracy, even compared to a naive automated approach that does not take domain expertise into account. However, we do not argue that domain experts should replace AI experts in representation debiasing. Rather, in line with previous studies highlighting the benefits and necessity of incorporating domain knowledge in AI development, fine-tuning, and evaluation~\cite{JungCSCW22, Dash2022, Sirocchi2024}, we contend that domain experts should be integral to the debiasing process alongside AI experts to mitigate limits of prediction models effectively.

\subsubsection{Partnership Between Domain Experts and AI Experts Is Crucial for Representation Debiasing}
While the benefits of involving domain experts in data augmentation and validation of generated data are clear, their interactions with AI systems can inadvertently introduce user interaction biases \cite{mehrabi2022survey}.  Our proposed \textsc{User-Interaction Bias Awareness} guideline acts as a warning to reduce the occurrence of user-interaction biases, but it cannot fully eliminate this bias.  In practice, we strongly recommend establishing a partnership between domain experts and AI experts to prevent harmful interactions within the AI system. More specifically, our study shows that domain experts with higher experience levels, deeper knowledge of different types of predictor variables used by the AI model, and a robust mental model of representation bias are the most effective when integrated into the debiasing process. Additionally, as proposed in prior works \cite{bhattacharya2024exmos, BhattacharyaUMAP24Demo},  rather than engaging individual domain experts, group consultations with domain experts of varied job roles and experience levels should be incorporated into the debiasing process. In general, our findings emphasise the need for a complementary role \cite{ZhangCHI20202} of domain experts and AI experts in representation debiasing.

\subsubsection{Recommendations to Enhance Understanding of Representation Bias}
As observed in our findings, a robust understanding of representation bias is essential for domain experts to participate in the debiasing process successfully.  However, our study raises an intriguing open research question: ``\textit{How can the understandability of representation bias be increased for domain experts?}'' Although a detailed answer to this research question would be of particular interest to the XAI community, the qualitative responses from our study participants offer valuable insights. Understanding representation bias and its impact on models can be challenging for many domain experts, and traditional visual explanations may have limitations in fully conveying this concept.  We suggest that Large Language Model (LLM)-based conversational chatbots could address this gap by supplementing the visual explanations \cite{lakkaraju2022rethinking, Slack2023}.  As highlighted in prior work by Miller \cite{Miller2017}, explanations through a natural conversation can offer a step-by-step approach to understanding concepts relevant to AI systems, such as the impacts of representation bias on the AI. Additionally, our study participants indicated a need for a query system where they can inquire about bias in specific patient groups (dataset segments) through question-answers. We envision that such chatbots could effectively explain representation bias and its implications through relevant examples.

\subsubsection{Combining Explanatory Debiasing with Explanatory Model Steering}
The concept of Explanatory Model Steering (EXMOS) \cite{bhattacharya2024exmos, BhattacharyaUMAP24Demo, bhattacharya2023_technical_report} explores the idea of leveraging XAI methods to engage domain experts in fine-tuning prediction models through their interactions with the training data. Although the goal of this research was to explore if domain experts can be successfully included in representation debiasing, their interactions during this process also serve to steer the model towards more unbiased and accurate predictions.  Therefore, integrating EXMOS interactions into our proposed debiasing system could be particularly beneficial. Incorporating components of a typical EXMOS system, such as manual and automated steering controls, can further facilitate domain experts to refine the resultant dataset that is obtained by combining newly generated data with the original training data. Moreover, an overall explanation dashboard for the model's training data could assist them in identifying and removing correlated features or other extreme values in the resultant dataset. Therefore, we believe that after retraining the model with the newly generated data, an EXMOS process will further enable effective collaboration between domain experts and AI experts, leading to a more streamlined and debiased AI system.

\subsubsection{Broader Applicability of the Design Guidelines}

To ensure the validity of our design guidelines, we prioritised highly cited and influential works alongside recent publications addressing emerging challenges in involving domain experts in AI fine-tuning (summarised in \Cref{tab:principle_methods}). Additionally, feedback from five healthcare experts during an exploratory session (\Cref{sec_exploratory_study}) informed initial refinements, followed by system evaluation with 35 experts to test the applicability of these guidelines. While these guidelines provide a concrete foundation for engaging domain experts in debiasing, they remain adaptable and open to refinement through broader expert involvement across diverse domains. Future work could enhance their generalisability by involving different types of stakeholders along with domain experts and AI experts, thereby promoting fairer and more transparent AI systems. Researchers can use these guidelines as a flexible framework for developing user-centred, debiased AI, with ongoing refinement ensuring their continued relevance and impact.

\subsection{Future Work}
Building on the improvements discussed, future research should focus on enhancing the understandability of representation bias for domain experts. Investigating methods to better communicate and visualise this bias could significantly improve their engagement in the debiasing process. Additionally, while our proposed guidelines have been applied within the healthcare domain, it is crucial to explore their relevance and effectiveness in other fields, such as finance and regulatory domains, where domain expertise plays a critical role. Moreover, evaluating our guidelines with unstructured datasets, such as images and text, is essential. Although the generic guidelines we propose are likely applicable to AI models trained on unstructured data, further user studies and experiments are necessary to validate their effectiveness and extendibility across diverse data types.

\section{Conclusion}
This paper introduces a set of generic design guidelines for effectively involving domain experts in the representation debiasing process. We demonstrated how these guidelines were implemented in a healthcare-specific debiasing application and validated them through an extensive mixed-methods user study with 35 healthcare experts.  Our findings confirm that domain experts can be successfully involved in the debiasing process by leveraging our proposed guidelines. Moreover, our work highlights the crucial role of domain experts in complementing AI experts during the debiasing process. Additionally, based on the insights generated from our study, we share certain recommendations for future researchers for effectively engaging domain experts in mitigating the impacts of representation bias.

\begin{acks}
We would like to express our gratitude Ivania Donoso-Guzmán, Maxwell Szymanski, Reza Samini for providing helpful suggestions that improved this work. We also extend our gratitude to the members of the Faculty of Health Science, University of Maribor, Slovenia, for helping us with the exploratory study. This research was supported by the Flanders AI Research Program (FAIR), KU Leuven Internal Funds (grant C14/21/072) and Research Foundation–Flanders (FWO grants G0A4923N and G067721N)~\cite{BhattacharyaCHIDC, Bhattacharya2024HowGoodIsYourExplanation}
\end{acks}
\bibliographystyle{ACM-Reference-Format}
\bibliography{references}


\end{document}